\documentclass[12pt,a4paper]{iopart}
\usepackage{amssymb,graphicx}
\usepackage{iopams}
\usepackage{setstack,cite}

\newcommand{\bea}{\begin{eqnarray}}
\newcommand{\eea}{\end{eqnarray}}
\newcommand{\bes}{\numparts}
\newcommand{\ees}{\endnumparts}

\begin{document}

\title[Non-autonomous BD solitons and Rabi oscillations in multi-component BECs]{Non-autonomous bright-dark solitons and Rabi oscillations in multi-component Bose-Einstein condensates}
\author{T Kanna$^1$, R Babu Mareeswaran$^1$, F Tsitoura$^2$, H E Nistazakis$^2$ and D J Frantzeskakis$^2$}
\address{$^1$Post Graduate and Research Department of Physics, Bishop Heber College, Tiruchirapalli--620 017, Tamil Nadu, India}
\address{$^2$Department of Physics, University of Athens, Panepistimiopolis, Zografos, Athens 15784, Greece}
\eads{kanna\_phy@bhc.edu.in (corresponding author), dfrantz@phys.uoa.gr}

\begin{abstract}
We study the dynamics of non-autonomous bright-dark matter-wave solitons in two- and three- component
Bose-Einstein condensates. Our setting includes a time-dependent parabolic potential and scattering
length, as well as Rabi coupling of the separate hyperfine states. By means of a similarity
transformation, we transform the non-autonomous coupled Gross-Pitaevskii equations into the
completely integrable Manakov model with defocusing nonlinearity, and construct the explicit form of
the non-autonomous soliton solutions. The propagation characteristics for the one-soliton state, and
collision scenarios for multiple soliton states are discussed in detail for two types of
time-dependent nonlinearities: a kink-like one and a periodically modulated one, with appropriate
time-dependence of the trapping potential. We find that in the two-component condensates the nature
of soliton propagation is determined predominantly by the nature of the nonlinearity, as well as the
temporal modulation of the harmonic potential; switching in this setting is essentially due to Rabi
coupling. We also perform direct numerical simulation of the non-autonomous two-component coupled Gross-Pitaevskii
equations to corroborate our analytical predictions. More interestingly, in the case of the
three-component condensates, we find that the solitons can lead to collision-induced energy
switching (energy-sharing collision), that can be profitably used to control Rabi switching
or vice-versa. An interesting possibility of reversal of the nature of the constituent soliton, i.e.,
bright (dark) into dark (bright) due to Rabi coupling is demonstrated in the three-component setting.
\end{abstract}
Journal Reference: {\it J. Phys. A: Math. Theor.} {\bf 46} (2013) 475201
\maketitle

\section{Introduction}
Studies on atomic Bose-Einstein condensates (BECs) have received considerable attention in recent
years \cite{Pethick}. In this context, it has been shown that nonlinear effects in matter waves can
emerge, with this particular direction having attracted much interest both in theory and in
experiments \cite{Pana1,Pana2}. From a theoretical viewpoint, this interest arises from the fact that
many of such nonlinear effects can be understood in the framework of lowest-order mean-field theory,
namely by means of the so-called Gross-Pitaevskii equation (GPE), which is nothing but the ubiquitous
nonlinear Schr{\"o}dinger (NLS) equation with external potential \cite{Pethick,Pana1,Pana2}.
The GPE is a nonlinear evolution equation (with the nonlinearity originating from the
interatomic interactions) and, as such, it permits the study of a variety of interesting purely
nonlinear phenomena. The latter, have primarily been studied by treating the condensate as a purely
nonlinear coherent matter-wave, i.e., from the viewpoint of the dynamics of nonlinear waves. Relevant
studies have already been summarized in various books \cite{Pana1} and reviews; see, e.g., \cite{Abd}
for bright solitons, \cite{DJ} for dark solitons, \cite{Fet} for vortices, \cite{MPLB} for dynamical
instabilities in BECs, and so on.

The formation and dynamical properties of solitons in BECs are determined by the nature of their
two-body atomic interactions, i.e., the sign of the $s$-wave scattering length which may be positive
(negative) for repulsive (attractive) inter-atomic interactions. For instance,
atomic dark (bright) solitons are formed in condensates with repulsive (attractive) interactions
\cite{Pana1,Pana2,Abd,DJ} (note that bright gap solitons are also possible in repulsive BECs, but
those are critically formed due to the presence of a periodic---optical lattice---potential
\cite{MKO}). On the other hand, it should be stressed that the above conditions can dramatically
change in the case of {\it multi-component} BECs, composed by two or more different hyperfine states
of the same atom species (e.g.,~$^{87}$Rb) \cite{{M.R},David}: in such a case, e.g., binary
condensates with repulsive inter- and intra-species interactions can support mixed bright-dark (BD)
solitons. This type of vector soliton consists of a dark soliton in one component coupled to a bright
soliton in the second component. These solitons are usually referred to as ``symbiotic'' solitons,
because the bright component cannot be supported in a stand-alone fashion (it is only supported as
such in attractive BECs \cite{Abd}; see also \cite{DJ}), and is only sustained because of the
presence of its dark counterpart, which acts as an effective external trapping potential. Such atomic
BD solitons were predicted in theory \cite{r23} and observed in experiments from different
experimental groups \cite{hamburg,engels1,engels2}. Notice that solitons in
three-component condensates (e.g., spinor $F=1$ BECs) have been studied too \cite{{J. I},r1,{L. Li}}
and mixed, BD solitons were predicted to occur as well \cite{bdspinor}.

On the other hand, coming back to the role of the sign and magnitude of the scattering length, it is
important to note that they can be adjusted over a relatively large range by employing external
magnetic, electric or optical fields near Feshbach resonances \cite{{E. A. Donley}}. This possibility
has given rise to many theoretical and experimental studies, with a prominent example being
the formation of bright matter-wave solitons and soliton trains in attractive condensates
\cite{{L. K}}, by switching the interatomic interactions from repulsive to attractive. Many
theoretical works studied the BEC dynamics under temporal modulation of the nonlinearity. In
particular, the application of such a ``Feshbach resonance management'' (FRM) technique
\cite{{P. G. Kev}} can be used to stabilize attractive higher-dimensional BEC against collapse
\cite{frm1}, to create robust matter-wave breathers in the effectively one-dimensional (1D)
condensate \cite{{P. G. Kev},frm2}, or compress bright solitons in the presence of expulsive
potentials \cite{liang}. Mathematically speaking, if the scattering length is time dependent then so
does the nonlinearity coefficient in the GPE, and the latter becomes {\it non-autonomous}. In
the pioneering works \cite{serkin}, it has been shown that the single-component non-autonomous GPE
can be transformed into the standard integrable NLS equation. The procedure developed in
Ref.~\cite{serkin} can be generalized to two-component condensates \cite{rajend} and the
corresponding non-autonomous matter-wave solitons were studied extensively with aid of the
explicit soliton solutions reported in Refs.~\cite{{Shepp},{T. K},{r43}}.

This work deals with the dynamics of multi-component BECs, particularly two- and
three-component BECs, with temporally modulated scattering length and trapping potential,
and in the presence of Rabi coupling; the latter is accounted for by a linear coupling between
separate wave functions induced by a radio frequency \cite{{M.R}} (see also
Refs.~\cite{{B. D},{I. M},{r35}} for a theoretical analysis and applications). Recently, coherent
many body Rabi oscillations have been observed experimentally and theoretically discussed in
Ref.~\cite{dudin}. Using two successive transformations, with one constraint equation being in the
form of a Ricatti equation, we reduce the original multicomponent non-autonomous GPEs to the
integrable defocusing vector NLS equation, known as the Manakov model \cite{manakov}. These
transformations can be viewed as a rotation followed by a similarity transformation.
This way, we find non-autonomous mixed (BD) solitons and analyze their dynamics and interactions in
detail, in both two- and three-component settings. We find that in the two-component system,
soliton propagation is determined predominantly by the nature of the nonlinearity, as well as
the temporal modulation of the harmonic potential; switching in this setting is essentially due to
Rabi coupling. Our analytical predictions are corroborated with results from direct simulations; very
good agreement between the two is found. Then by going one step further in the case of the three-component
system, we find that the solitons can lead to collision-induced energy switching (energy-sharing
collision) that can be profitably used to control Rabi switching or vice-versa. An interesting
possibility of reversal of the nature of the constituent soliton, i.e., bright (dark) into dark
(bright) induced by the Rabi coupling is demonstrated in the three-component setting.

The remaining part of this paper is arranged as follows. In Sec. II, we introduce the generalized
non-autonomous coupled GPEs, for two-component condensates, with time-varying interaction
strength and Rabi coupling. Also, we construct the one- and two-soliton solutions to the 1D coupled
non-autonomous GPEs and analyze their propagation characteristics. In Sec. III, we
consider the three-component non-autonomous GPEs with Rabi coupling, and explore the
interesting energy exchange collisional features of non-autonomous matter-wave solitons. Finally, in
Sec. IV, we present our conclusions.

\section{Non-autonomous BD solitons in two-component BECs and Rabi oscillations}

In the mean field approximation, the dynamics of two-component condensates can be well described by a
pair of two coupled three-dimensional (3D) GPEs (see, e.g., Ref.~\cite{williams}). In the
physically relevant case of highly anisotropic (cigar-shaped) traps, these GPEs can be
reduced to an effectively 1D system (see, e.g., Ref.~\cite{Pana2}), which is of the following
dimensionless form:
\bea
\hspace{-0.5cm}i \psi_{j,t}=-\frac{1}{2}\psi_{j,xx}+ \sum_{l=1}^2 g_{jl}(t)|\psi_l|^2\psi_j+\sum_{l=1,(l\neq j)}^2\sigma_l\psi_l+ V_{j}(x,t)\psi_j,\quad j=1,2.
\label{twogp}
\eea
Here, $\psi_j$ is the macroscopic wave function of the $j^{th}$ component, time $t$ and spatial
coordinate $x$ are respectively measured in units of the inverse transversal frequency
$\omega_\bot^{-1}$ and the transverse harmonic oscillator length $a_0=\sqrt{\hbar/m\omega_\bot}$
($m$ denotes atomic mass). In Eq.~(\ref{twogp}) the nonlinearity coefficients $g_{jl}$ are
proportional to the $s$-wave scattering lengths $a_{jl}$, namely
$g_{jl}(t)=2 a_{jl}(t)/a_B$ (where $a_B$ is the Bohr radius), and are assumed to be subject to the
FRM technique (see below); furthermore $V_j(x,t)$ are the external potentials
[for $V(x,t) < 0$~($V(x,t) > 0$) the potential is expulsive (confining)], and the linear cross
coupling parameters $\sigma_{\it l}$ represent the Rabi coupling: in fact these coefficients are
proportional to the Rabi frequency, and are responsible for the transfer of atoms of component
$\psi_1$ to component $\psi_2$ and vice-versa. In our case, considering two different hyperfine
states of the same atom, the external time-varying potentials $V_j$ are equal, i.e.,
$V_1=V_2=V = (1/2)\Omega^2(t)x^2$. Here, the strength of the parabolic trap is given by
$\Omega^2(t) = \omega_x^2(t)/\omega_\bot^2$, where $\omega_x(t)$ is the temporally modulated axial
trap frequency. Finally, the macroscopic wave functions $\psi_j$ are normalized such that
$\int_{-\infty}^{\infty}|\psi_j|^2 dx=N_j$ ($j= 1,2$), where $N_j$ is the number of atoms in the
${\it j}^{th}$ component.

\subsection{Non-autonomous BD one-soliton solution}
\indent Based on experimental results pertaining to two-component $^{87}$Rb BECs (see, e.g.,
\cite{engels1}), we can assume that the scattering length ratios are almost equal to one,
and also they can be tuned through Feshbach resonance (as it was demonstrated in the experimental
works \cite{ingu,papp}). Hence,
we consider the case of equal interaction strengths,
i.e., $g_{jl}=\rho(t)$ in Eq.~(1), and choose
the linear coupling coefficients $\sigma_l$ 
as $\sigma_1=\sigma_2=\sigma$, with $\sigma>0$.

In such a case, as shown in Appendix~A, we can use two successive transformations to map
the non-autonomous system (\ref{twogp}) into an integrable autonomous nonlinear system.
This is possible mainly due to the temporal dependence
of the nonlinearity and external harmonic potential; in the absence of such dependence these
transformations are not possible, in general. Notice, however, that in the absence of potential
(homogeneous system), one can transform Eq.~(1) into an integrable coupled nonlinear Schr\"odinger (CNLS) system even in the
absence of inhomogeneities. The above mentioned transformations are:
(i) a unitary transformation
(see, e.g., Refs.~\cite{{B. D},r35}) to reduce Eq.~(1) to a non-autonomous system of equations
without the Rabi coupling term;
(ii) a similarity transformation, which reduces the aforementioned system
to the following integrable defocusing CNLS equations \cite{Radha,proceedmv}:
\bea
&iq_{j,T}+ q_{j,XX}- 2 \displaystyle\sum_{l=1}^2|q_l|^2q_j=0,\quad j=1,2.
\label{2manakov}
\eea
%
Equation~(\ref{2manakov}) is the completely integrable Manakov system and admits $N$-soliton
solutions of dark-dark and mixed (BD) types.
Here, we focus only on mixed type soliton solutions due to their recent experimental observations
and for their special dynamical features \cite{hamburg,engels1}.
%


The mixed one-soliton solution of Eq.~(\ref{2manakov}) can be obtained by Hirota's direct
method \cite{Hirota}. In standard form, the mixed one-soliton solution of the two-component
defocusing CNLS equations with the bright (dark) part appearing in the $q_1$ ($q_2$) component
(see Refs.~\cite{Shepp,proceedmv}) is given in Appendix B.1. Using the analytical form of this type of
soliton, as well as the transformations of Appendix~A, we find an exact analytical
non-autonomous soliton solution of Eq.~(\ref{twogp}) in the form:
\begin{eqnarray}
\psi_1 = \mbox{cos}(\sigma t)\phi_1- i\sin(\sigma t)\phi_2, \qquad
\psi_2 = \mbox{cos}(\sigma t)\phi_2- i\sin(\sigma t)\phi_1, \label{ps1}
\end{eqnarray}
where $\phi_1$ and $\phi_2$ are given by:
\bea
\label{N-one}
\!\!\!\!\!\!\!\!\!\!\!\!\!\!\!\!\!\!\!\!\!\!\!\!\!\!\!\!\!\!\!\!\!\!\!\!\!\!\!\!
&&\phi_{1}(x,t) =  \xi_1 \sqrt{2\rho(|c_{1}|^2 \mbox{cos}^2 \varphi_{1}- k^2_{1R})}~e^{i(\bar{\eta}_{1I}+ \theta+\tilde{\theta})} \nonumber\\
\!\!\!\!\!\!\!\!\!\!\!\!\!\!\!\!\!\!\!\!\!\!\!\!\!\!\!\!\!\!\!\!\!\!\!\!\!\!\!\!
&&\times  \mbox{sech}\left(k_{1R}(\sqrt {2}~\xi_1 (\rho x - 2\xi_2 \xi_1^2\int_0^t \rho^2 dt)- 2k_{1I}\xi_1^2 \int_0^t \rho^2 dt)+R/2\right), \\
\hspace{-2.5cm}
&&\phi_{2}(x,t) = -c_{1}\xi_1 \sqrt{2\rho}~e^{i(\bar{{\zeta_1}}+\varphi_1+\tilde{\theta})} \nonumber\\
\hspace{-2.5cm}
&&\times \left(\mbox{cos}\varphi_1~\mbox{tanh}[k_{1R}(\sqrt {2}~\xi_1 (\rho x - 2\xi_2 \xi_1^2\int_0^t \rho^2 dt)-2k_{1I}\xi_1^2 \int_0^t \rho^2 dt)+R/2]+ i~ \mbox{sin}\varphi_1\right),
\eea
where $\bar{\eta}_{1I} = k_{1I}\sqrt{2}\xi_1(\rho x-2\xi_2\xi_1^2\int_0^t\rho^2dt)-(k_{1R}^2-k_{1I}^2- 2|
c_1|^2) \xi_1^2\int_0^t\rho^2 dt$, $\bar{\zeta_1} = -(b_1^2 + 2|c_1|^2)\xi_1^2 \int_0^t\rho^2 dt + b_1 \sqrt {2}~\xi_1 (\rho x - 2\xi_2 \xi_1^2\int_0^t\rho^2 dt)$
and the other parameters and functions involved in the above equations are
defined in Appendices A and B.
%

The above mixed soliton bears resemblance to the recently observed, so-called ``beating'' dark-dark
soliton \cite{engels2,pe5}, where the bright and dark parts coexist in the same component with a
constant asymptotic value as $|x|\rightarrow \infty$; however, the form of this BD soliton still
depends on the Rabi coupling and the time-dependent nonlinearity and parabolic potential. Notice that
due to the presence of Rabi coupling, there will be a periodic switching between the two components
along with oscillating background. Another important observation for the integrable two-component
case is the dependence of
the existence of e.g., the bright part of the
mixed soliton on its dark counterpart (see Eqs.~(\ref{1solXT1})-(\ref{1solXT2}) and relevant
discussion in Appendix B.1). Thus,
it is impossible to make
any one of the constituents (bright/dark) of the mixed soliton to vanish completely. In fact, this
will result in singularities in the other soliton part. This holds even in the presence of the Rabi
coupling for the two-component system.

To elucidate the dynamics of such special type of
mixed soliton we consider the following two examples.

\noindent\underline{Example 1:} {\bf{Kink-like nonlinearity}}
\\
\indent First we note that a fast time modulated nonlinearity displays interesting dynamics in BECs \cite{Abdu}. A relatively sudden jump in the nonlinearity coefficient can be well represented by a kink-like nonlinearity of the form:
\bes\bea
\rho(t) = 1 + \mbox{tanh}(\omega t + \delta),
\label{kink}
\eea
with the associated atomic scattering length being $a_s(t)= \frac{1}{2}a_B[1 + \mbox{tanh}(\omega t + \delta)]$; here, $\omega$ denotes the time scale characterizing the jump, and $\delta$ is an arbitrary constant.

\begin{figure}[tbp]
\centering\includegraphics[width=0.44\linewidth]{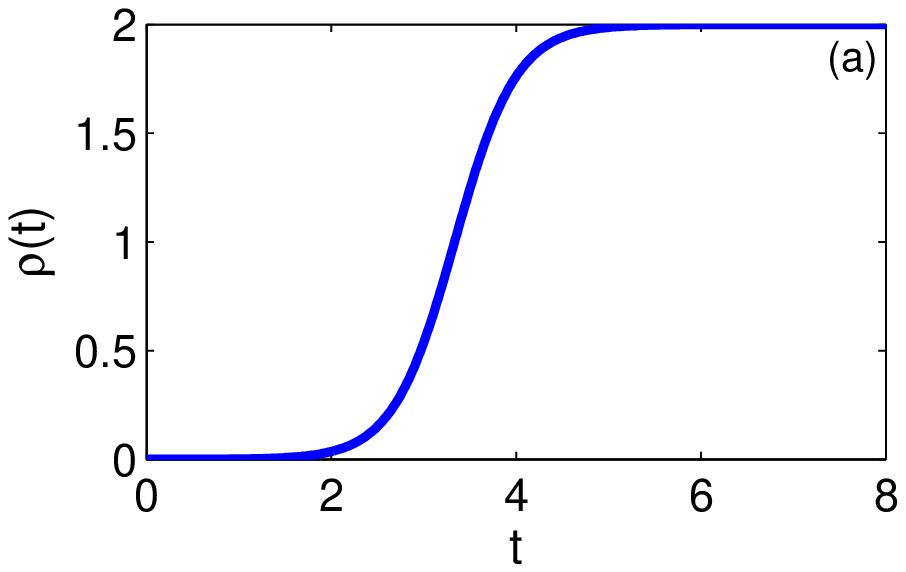}~~~\includegraphics[width=0.44\linewidth]{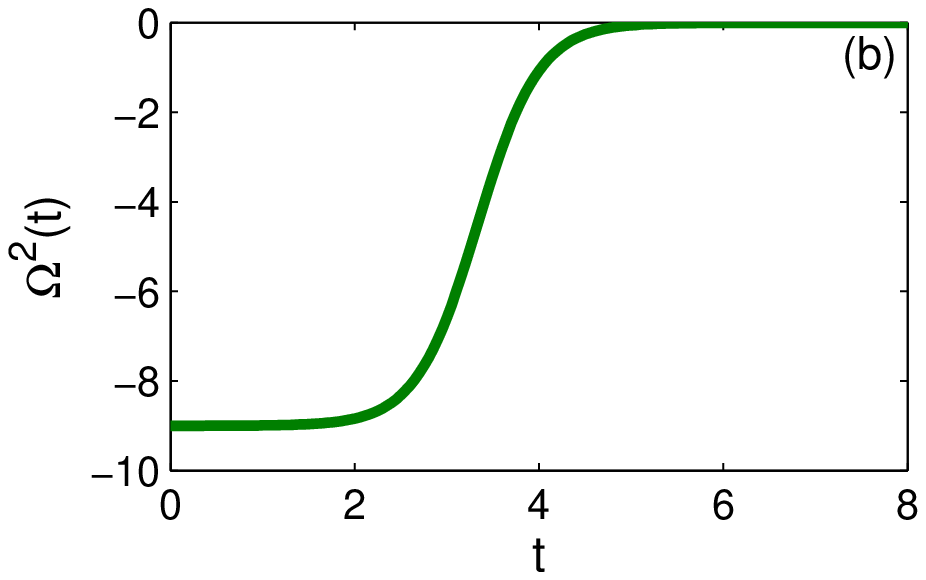}
\caption{Typical form of the kink-like nonlinearity $\rho$(t) (a) and corresponding strength
of external harmonic potential (b).}
\label{fig1}
\end{figure}
This is shown in Fig.~\ref{fig1}(a) and the corresponding external harmonic potential is displayed in Fig.~\ref{fig1}(b). The time-dependent nonlinearity $\rho$(t) admitting the above form can be realized for the following scattering length \cite{serkin},
\bea
&&\frac{a_s(t)}{a_{bg}}=\left(1+\frac{\Delta}{B_0-B(t)}\right),
\label{Feshformula}
\eea
where $a_{bg}$ is the value of scattering length far from the Feshbach resonance, $B(t)$ is the
applied time-varying magnetic field, $B_0$ is the resonant value of the magnetic field, and $\Delta$
is the resonance width in the presence of the magnetic field. The corresponding strength of the time-
dependent magnetic trap is determined by a Ricatti equation [see Eq.~(\ref{ricatti})
in Appendix~A] as follows:
\bea
\Omega ^2(t) =-\left(\frac{4\omega^2}{1+e^{2(\omega t+\delta)}}\right).
\label{potstrength}
\eea\ees

The propagation of the non-autonomous mixed soliton in the absence and presence of Rabi coupling are depicted in the top panels of Figs.~2 and 3, respectively. We have also solved the system (\ref{twogp}) numerically by means of the split-step Fourier method, and the resulting numerical plots for the above choice of the time-dependent nonlinearity and potential strength (\ref{potstrength}) are given in the bottom panel.

\begin{figure}[h]
\centering\includegraphics[width=0.9\linewidth]{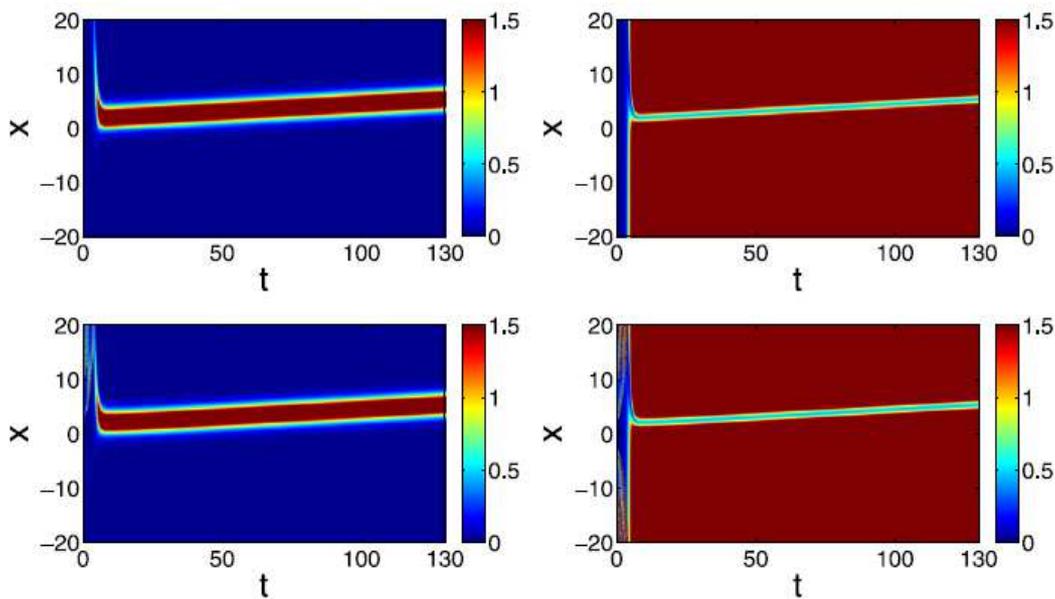}
\caption{Two-component one-soliton solution of GPEs.~(\ref{twogp}) with kink-like nonlinearity in the absence of Rabi coupling. Top panels show the intensity plots of exact analytical results (upper left panel: bright component, and right upper panel: dark component of the mixed soliton), while bottom panels show numerical results (left bottom panel: bright component, and right bottom panel: dark component of the mixed soliton). The parameter values are $k_1 = 0.5 + 0.02i$, $b_1$ = 0.2, $c_1$ = 2, $\xi_1$ = 0.5, $\xi_2$ = 0, $\alpha_1^{(1)}$ = 0.5, $\delta = -4$ and $\omega = 0.7$.}
\end{figure}

Comparing the top and bottom panels, one can see that there is a very good agreement between numerical and analytical results. We find that, in the absence of the Rabi coupling, there are no oscillations for the solitons and the background. However, the shape, width and velocity of solitons are modulated. Then, the introduction of Rabi term leads to an exchange of number of atoms between the components, which ultimately results in beating oscillations, as shown in Fig.~3. For small values of $\sigma$, the background just starts to oscillate (see Fig.~3). However, by increasing the parameter $\sigma$, one can observe rapid oscillations of the background also. To illustrate this, the contour plots of $|\psi_1|^2$ for $\sigma = 0.1$ and $\sigma = 0.2$ are shown in Fig.~4. Similar observations can be made for the second component, $(\psi_2)$, as well. These oscillations of the background are due to the exchange of condensates between the soliton and the background.

\begin{figure}[h]
\centering\includegraphics[width=0.9\linewidth]{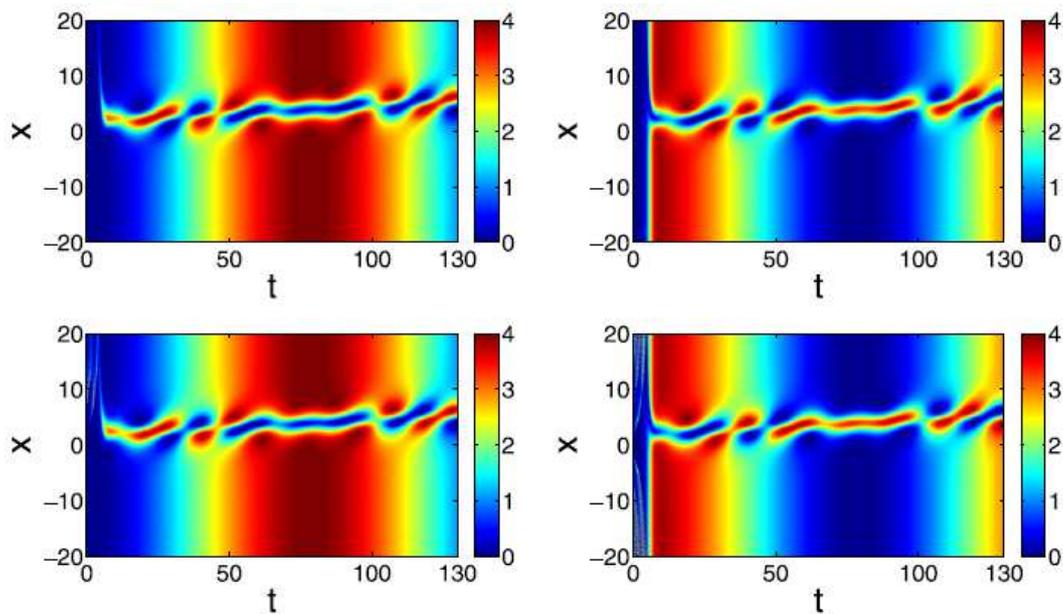}
\caption{Two-component mixed one-soliton solution with co-existing bright-dark parts for the kink-like nonlinearity in the presence of Rabi coupling: Analytical results (upper left panel: $\psi_1$ component, and right upper panel: $\psi_2$ component of the mixed soliton). Numerical results (left bottom panel: $\psi_1$ component, and right bottom panel: $\psi_2$ component of the mixed soliton).
The parameter values are $k_1 = 0.5 + 0.02i$, $b_1$ = 0.2, $c_1$=2, $\xi_1$ = 0.5, $\xi_2$ = 0, $\alpha_1^{(1)}$ = 0.5, $\delta$ = -4, $\omega$ = 0.7 and $\sigma$ = 0.02.}
\end{figure}

Another important observation following from Figs.~3 and 4, as well as from expressions (\ref{ps1})-(5) is that the increase of the value of Rabi coupling leads a significant portion of the dark part of mixed soliton to appear in the first component ($\psi_1$), along with the bright part accompanied by beating effects. Similar effects take place in component $\psi_2$ too. Thus, in a given component one can have co-existing oscillating BD soliton.
\\
\begin{figure}[h]
\centering\includegraphics[width=0.9\linewidth]{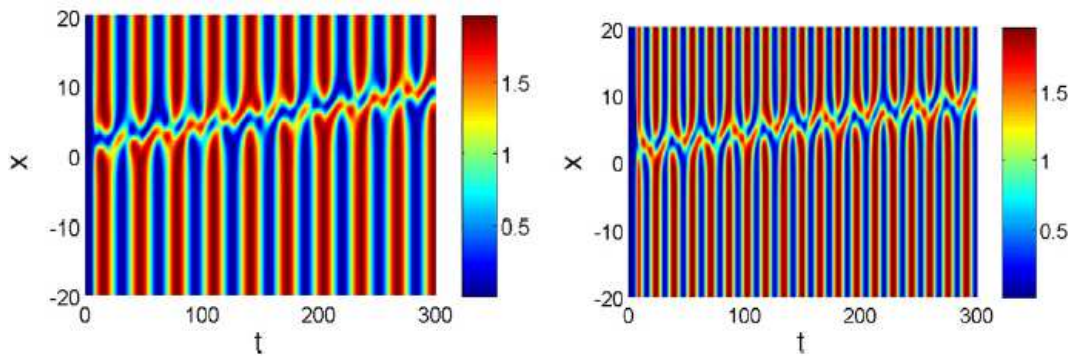}
\caption{Mixed one-soliton solution for the kink-like nonlinearity in left panel for $\sigma = 0.1$ and right panel for $\sigma = 0.2$. The other parameters are fixed as $k_1 = 0.5 + 0.02i$, $b_1 = 0.2$, $\xi_1 = 0.5$, $\xi_2 = 0$, $\alpha_1^{(1)} = 0.5$, $\delta = -4$ and $\omega = 0.7$.}
\end{figure}
Enlarging Fig.~3 in the region $t = 0$ to $5$, we observe that the non-autonomous mixed soliton is absent in this region. This is due to the form of the nonlinearity $\rho(t)$, which becomes zero in the range $t$ = $0$ to $5$ and reaches a saturation for large positive $t$ values. Thus, even though the Rabi coupling leads to an exchange of atoms among the two components leading to oscillations, the nature of soliton propagation is predominantly determined by the nature of  time-dependent nonlinearity.\\

\noindent\underline{Example 2:} {\bf{Periodically modulated nonlinearity}}\\
\indent Next, we consider another physically interesting example, namely the case of a periodic time-varying nonlinearity of the form:
\bes\bea
 \rho(t) = 1 +\varepsilon\mbox{cos}(\omega t + \delta),
 \label{periodic}
 \eea
where $\varepsilon$ is an arbitrary real constant, $\omega$ is the characteristic frequency, and $\delta$ is a real constant parameter. In this case, the form of atomic scattering length is $a_s(t)= \frac{1}{2}a_B[1 + \varepsilon\mbox{cos}(\omega t + \delta)]$. Figure 5(a) represents the
form of such a time-varying nonlinearity. The expression for the pertinent time-dependent external magnetic field $B(t)$ required to achieve the above form of the nonlinearity can be determined from the formula (\ref{Feshformula}). The corresponding strength of the magnetic trap admits the form
\bea
\Omega ^2(t) =\omega^2\varepsilon\left(\frac{-3\varepsilon-2 \mbox{cos}(\omega t+\delta)+\varepsilon\mbox{cos}[2(\omega t+\delta)]}{2[1+\varepsilon\mbox{cos}(\omega t+\delta)]^2}\right),
\label{perio_pot_str}
\eea\ees
in order to satisfy Eq.~(\ref{ricatti}). Note that the potential is sign reversible and can support the same type of soliton for both positive and negative signs; see Fig.~5(b).

\begin{figure}[h]
\centering\includegraphics[width=0.39\linewidth]{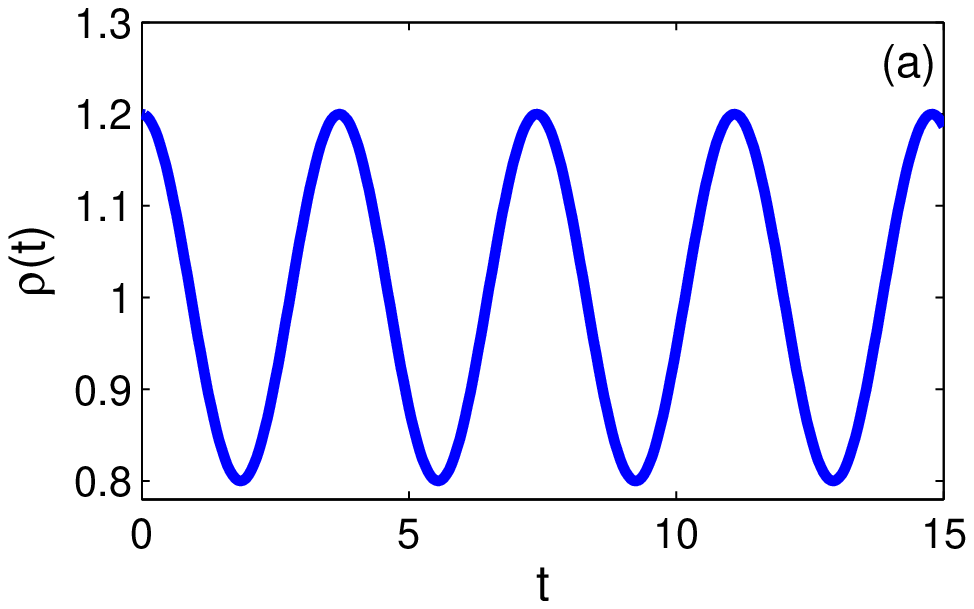}~~~~\includegraphics[width=0.39\linewidth]{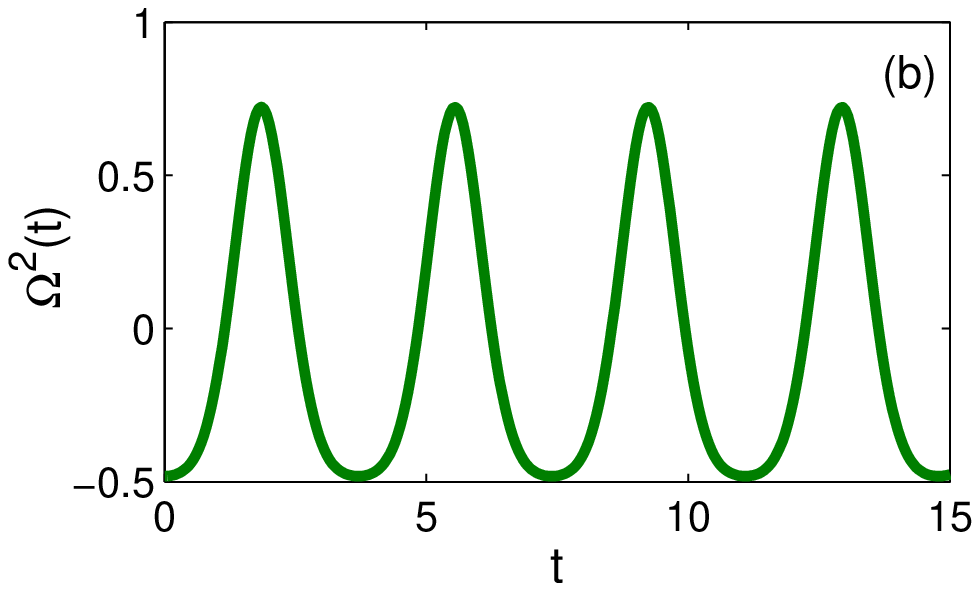}
\caption{Periodically modulated nonlinearity $\rho$(t) (a) and corresponding strength of the external harmonic potential (b).}
\end{figure}

It is apparent from the periodic nature of the nonlinearity, and the time modulation of the strength of the external potential, that the condensates also execute periodic oscillations in both the components even in the absence of linear coupling as shown in Fig.~6. The numerical results, obtained by a direct integration of Eq.~(\ref{twogp}) are shown in the bottom panel of Fig.~6 in the absence of Rabi
coupling.
\begin{figure}[h]
\centering\includegraphics[width=0.9\linewidth]{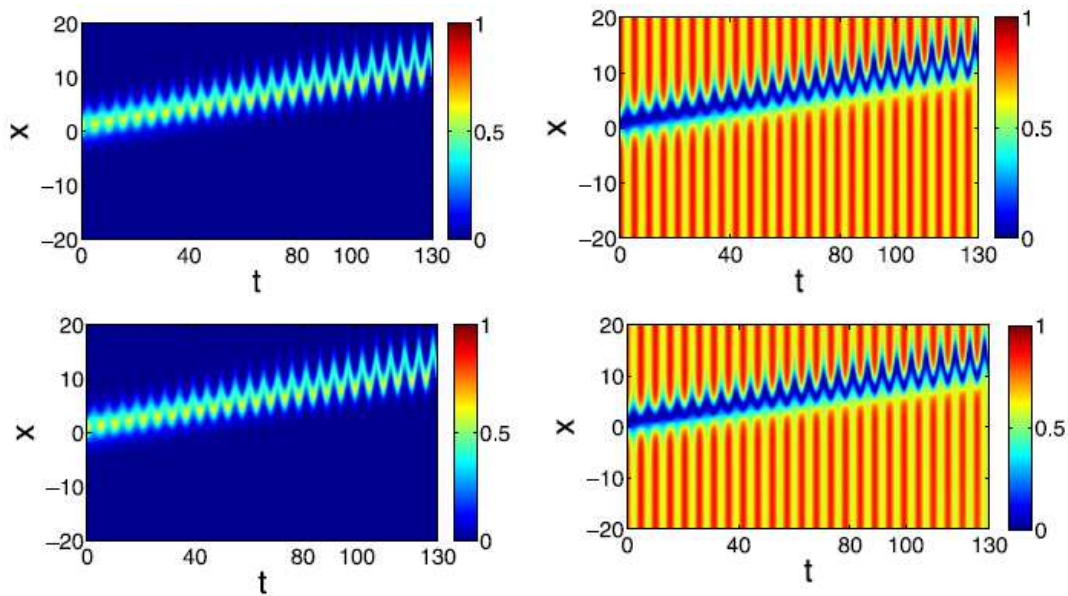}
\caption{Two component mixed one-soliton solution of Eq.~(1) for the periodically modulated nonlinearity in the absence of Rabi coupling. Top panels show the exact analytical results (upper left panel: bright component, and right upper panel: dark component) and bottom panels show numerical results (left bottom panel: bright component, and right bottom panel: dark component). The parameter values are $k_1 = 0.5 + 0.1i$, $b_1 = 0.2$, $\alpha_1^{(1)} = 1.5$, $c_1 = 1$, $\xi_1 = 0.6$, $\xi_2 = 0$, $\varepsilon = 0.2$, $\omega = 1.2$ and $\delta = 0$. }
\end{figure}
Next, we introduce the Rabi coupling between the two components; then, as expected, there will be an exchange of atoms between the components, as well as between the soliton and the background. This exchange induces oscillations in the density of condensates which, in turn, modulate the oscillations due to the periodic nature of $\rho(t)$ and $\Omega^2(t)$ and result in soliton beating in both components. This becomes clear in Fig.~7, from the plots showing the non-autonomous soliton (\ref{ps1}) for the choice (\ref{periodic}) (upper panel) and the corresponding numerical results (bottom panel).

\begin{figure}[h]
\centering\includegraphics[width=0.9\linewidth]{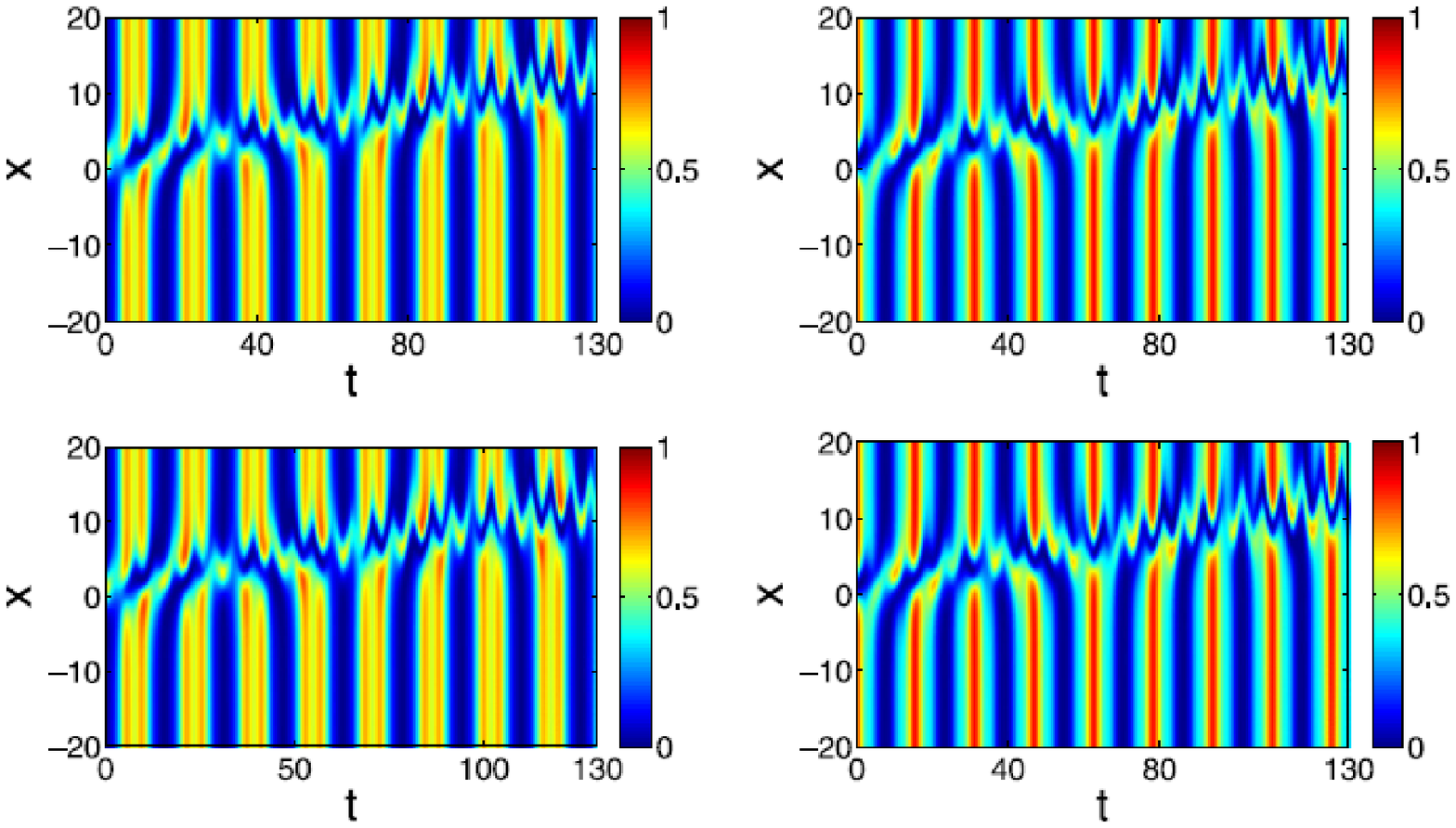}
\caption{Two component one-soliton solution of Eq.~(1) for the periodically modulated nonlinearity in the presence of Rabi switch. Top panels show the exact analytical results (upper left panel: $\psi_1$ component, and right upper panel: $\psi_2$ component) and bottom panels the numerical results (left bottom panel: $\psi_1$ component, and right bottom panel: $\psi_2$ component). The parameter values are $k_1 = 0.5 + 0.1i$ , $b_1 = 0.2$, $\alpha_1^{(1)} = 1.5$, $c_1 = 1$, $\xi_1 = 0.6$, $\xi_2 = 0$, $\varepsilon = 0.2$, $\omega = 1.2$, $\delta = 0$ and $\sigma = 0.2.$}
\end{figure}

As in the previous example, here also both parts of the mixed soliton co-exist in a given component due to Rabi switching.
The left panel of Fig.~8 displays the propagation of mixed soliton for a small value of $\sigma$ (=$0.1$). It is observed that the soliton only oscillates periodically, but the background is not oscillating. The right panel of Fig.~8 displays that for larger value of $\sigma$ (say $\sigma=0.2$), the background oscillates rapidly and significant switching of dark and bright parts among the components occurs, thereby resulting in the co-existence of both dark and bright parts in the same component. By increasing the values of $\omega$, we also observe ``creeping'' soliton propagation with beating effects, which are not presented here. Similar type of creeping soliton appears in the presence of inhomogeneities too.
\\

\begin{figure}[h]
\centering\includegraphics[width=0.9\linewidth]{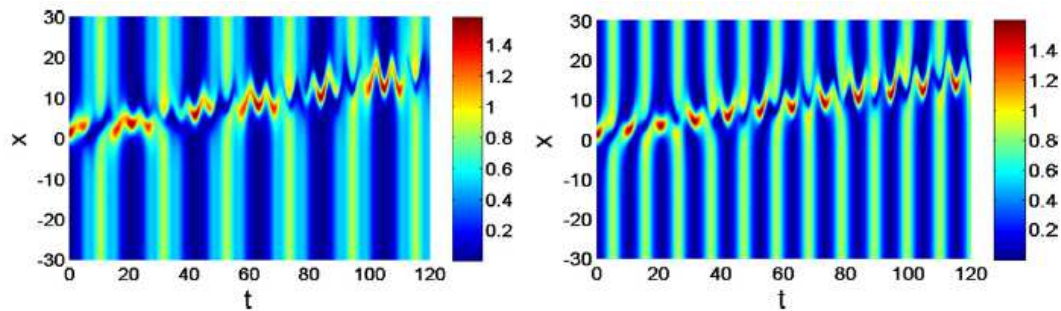}
\caption{Intensity plots of non-autonomous one-soliton solution in the $\psi_1$-component [see Eq.~7(a)] for the
periodically modulated nonlinearity for $\sigma = 0.1$ (left panel) and $\sigma = 0.2$ (right panel). The other parameters are fixed as $k_1 = 0.5 + 0.1i$ , $b_1 = 0.2$, $\alpha_1^{(1)}$ = 1.5, $c_1 = 1.5$ , $\xi_1 = 0.6$, $\xi_2 = 0$, $\varepsilon = 0.2$, $\omega = 1.2$ and $\delta = 0$.}
\end{figure}

\indent From the above two examples, we can conclude that in the non-autonomous two-component GPEs.~(\ref{twogp}) the nature of soliton propagation is determined predominantly by the temporal dependence of $\rho(t)$ and $\Omega^2(t)$  while the switching of condensates is completely dependent on the Rabi coupling.

\subsection{Two soliton solution of non-autonomous two-component GPEs}

We now turn our attention to the case of mixed two-soliton solution of the integrable
2-CNLS system (\ref{2manakov}). The explicit form of this solution, given in Appendix~B.1,
contains all the information regarding the collision of two solitons. We will employ this autonomous soliton solution
to construct the exact two soliton solution of the corresponding non-autonomous system (\ref{twogp}) in detail.

The asymptotic analysis of the two-soliton solution of the two component GPEs system (\ref{2manakov}) given in Appendix C.1, shows that the two solitons undergo standard elastic collision as $|A_i^{l+}|$=$|A_i^{l-}|$, $i,l=1,2$, where $A_i^l$ represents the amplitude of $l^{th}$ soliton in $i^{th}$ component. Here and in the following, the superscript (subscript) of A (or $\psi$) represents the number of soliton (component) while -(+) appearing in the corresponding quantities indicates their form before (after) collision. Thus the
amplitude and speed of the bright and dark components are preserved after the collision, except for a
phase-shift $(\Phi = \frac{R_3-R_2-R_1}{2})$, where $R_1, R_2$ and $R_3$ are defined in Appendix $B.1$. This elastic collision of solitons is shown in Fig.~9.

\begin{figure}[h]
\centering\includegraphics[width=0.9\linewidth]{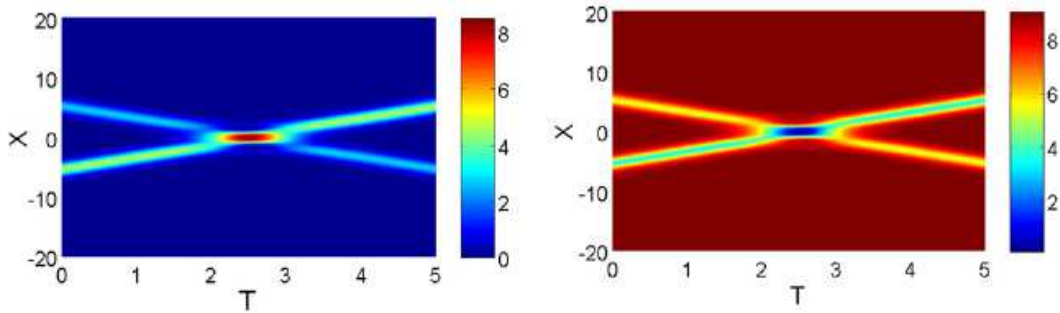}
\caption{Two soliton collision of system (\ref{2manakov}) [see Eq.~(\ref{2c2sol})]. The soliton parameters are fixed as $k_1 = -1 + i$, $k_2 = 1 - i$, $c_1 = 3$, $b_1= 0.2$ and $\alpha_1^{(1)} = \alpha_2^{(1)} =0.02$.}
\end{figure}

The exact dynamics of the non-autonomous two mixed solitons can be understood after
transforming the two-soliton solution [see Eqs.~(B.3)-(B.13)] by using the transformations given in Appendix~A
[see Eqs.~(\ref{similarity}) and (\ref{rabi})]. In the following analysis, the collision dynamics for two interesting forms of the  nonlinearity coefficient $\rho(t)$, namely
kink-like and the periodically modulated nonlinearity are considered in detail.

\subsection*{(i) Collision dynamics in the presence of kink-like modulated nonlinearity}

Figure~10 shows the two-soliton collision in the non-autonomous two-component GPEs (\ref{twogp}) with kink-like nonlinearity, whose form is given by Eq.~(\ref{kink}) in the absence of Rabi terms, and the corresponding strength of the magnetic trap is given by Eq.~(\ref{potstrength}). The figure shows that the two solitons undergo a collision around $t$ = $80$ and get well separated for larger values of $t$.
\begin{figure}[h]
\centering\includegraphics[width=0.9\linewidth]{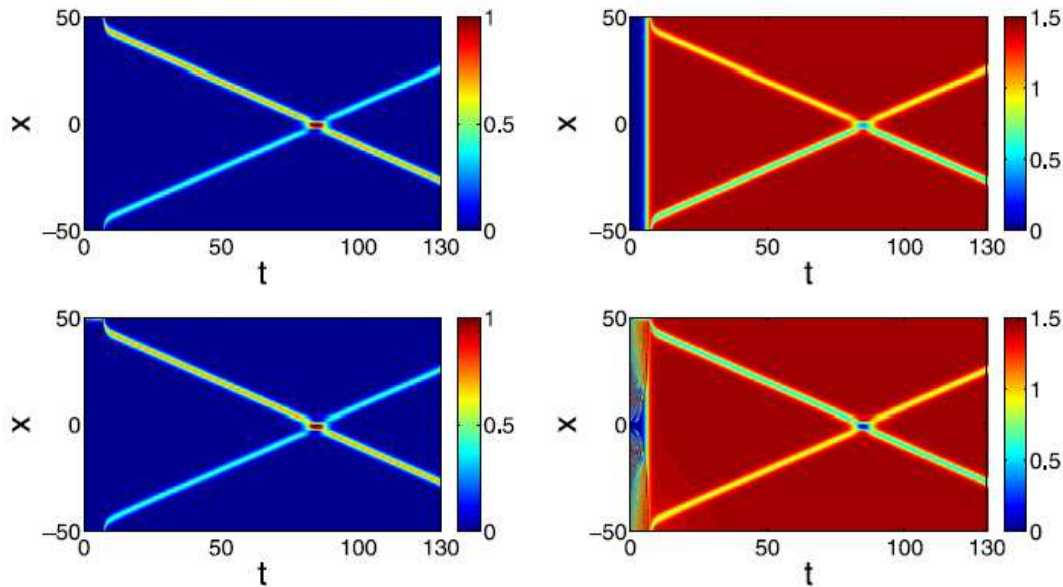}
\caption{Interaction of non-autonomous solitons in two-component GPEs for the kink-like nonlinearity
in the absence of Rabi coupling. Top panels show the analytical results (upper left panel: bright
component, and right upper panel: dark component) and bottom panels show the numerical results (left
bottom panel: bright component, and right bottom panel: dark component). The soliton parameters are
$k_1 = -1 + i$, $k_2 = 1 - i$, $c_1 = 3$, $b_1= 0.2$, $\alpha_1^{(1)} = \alpha_2^{(1)} = 0.03$,
$\delta= -4.5$, $\xi_1 = 0.2$, $\xi_2 = 0$ and  $\omega = 0.75$.}
\end{figure}
Next, we include the Rabi coupling and plot the non-autonomous two soliton collision in Fig.~11. We
notice from Figs.~10 and 11 that the nature of soliton collision in the presence of Rabi term looks
alike the soliton collision in the absence of Rabi term. Additionally, in the present case, there is
an oscillating exchange of atoms between the two components. The background oscillates in a periodic
manner, in which the oscillations in a given component is maximum while it is minimum in the other.

Here we present the asymptotic analysis of the non-autonomous two-soliton solution of two-component GPEs with kink-like nonlinearity, obtained by making use of (B.3)-(B.13) and the transformations (\ref{rabi}) and (\ref{similarity}).
For the choice $k_{1R}<k_{2R}$, $k_{1I}>k_{2I}$, the asymptotic forms of the non-autonomous solitons ($S_l$, $l=1,2$) well before and after collision can be expressed as below:\\
{\bf{Before collision}}
\bea
\psi_1^{l-}&= \xi_1\sqrt{2\rho}~~[A_1^{l-} \mbox{cos}(\sigma t)\mbox{sech}(\eta_{lR}+R_l/2)e^{i(\eta_{lI}+\tilde{\theta})}\nonumber\\
& - i A_2^{l-}\mbox{sin}(\sigma t)(\mbox{cos}\varphi_l\mbox{tanh}(\eta_{lR}+R_l/2)+i\mbox{sin}\varphi_l)],\label{kink-asy-2com1}\\
\psi_2^{l-}&= \xi_1\sqrt{2\rho}~~[A_2^{l-}\mbox{cos}(\sigma t)(\mbox{cos}\varphi_l\mbox{tanh}(\eta_{lR}+R_l/2)+i\mbox{sin}\varphi_l),\nonumber\\
& - i A_1^{l-}\mbox{sin}(\sigma t)\mbox{sech}(\eta_{lR}+R_l/2)e^{i(\eta_{lI}+\tilde{\theta})}], \quad l=1,2,
\label{kink-asy-2com2}
\eea

\noindent{\bf{After collision}}
\bea
\psi_1^{l+}&= \xi_1\sqrt{2\rho}~~[A_1^{l+} \mbox{cos}(\sigma t)\mbox{sech}(\eta_{lR}+(R_3-R_{3-l})/2)e^{i(\eta_{lI}+\tilde{\theta})}\nonumber\\
& - i A_2^{l+}\mbox{sin}(\sigma t)(\mbox{cos}\varphi_l\mbox{tanh}(\eta_{lR}+(R_3-R_{3-l})/2)+i\mbox{sin}\varphi_l)],\\
\psi_2^{l+}&= \xi_1\sqrt{2\rho}~~[A_2^{l+} \mbox{cos}(\sigma t)(\mbox{cos}\varphi_l\mbox{tanh}(\eta_{lR}+(R_3-R_{3-l})/2)+i\mbox{sin}\varphi_l)]\nonumber\\
& - i A_1^{l+}\mbox{sin}(\sigma t)\mbox{sech}(\eta_{lR}+(R_3-R_{3-l})/2)e^{i(\eta_{lI}+\tilde{\theta})}], \;\;l=1,2,
\eea
where $A_1^{l-} = \alpha_l^{(1)}e^{-\frac{R_l}{2}}$, $A_2^{l-} = -c_1e^{i(\zeta_1+\varphi_l+\tilde{\theta})}$, $A_1^{l+} = \frac{1}{2}e^{(2\delta_{l1}-R_3-R_{3-l})/2}$, $A_2^{l+}=c_1 e^{i(\zeta_1+2\varphi_{3-l}+\varphi_l+\tilde{\theta})}$, $\rho = 1+\mbox{tanh}(\omega t+\delta)$, $\eta_{lR} = k_{lR}(\sqrt{2}\xi_1(\rho x - 2\xi_2\xi_1^2\int_0^t \rho^2 dt)-2k_{lI}\xi_1^2\int_0^t\rho^2 dt)$ and $\eta_{lI}=k_{lI}(\rho x - 2\xi_2\xi_1^2\int_0^t \rho^2 dt)-(k_{lR}^2-k_{lI}^2-2|c_1|^2)\xi_1^2\int_0^t\rho^2 dt$,
where $\tilde{\theta}= -\frac{\rho_t}{2\rho}x^2+2\xi_2\xi_1^2(\rho x - \xi_2\xi_1^2\int_0^t \rho^2 dt)$, and all other quantities are given in Appendix B [cf.~(B.6)-(B.13)].
Here, $A_j^{l-}$ ($A_j^{l+}$) is the amplitude of the soliton $S_l$ in the $j$th component, $l,j=1,2$, before (after) collision, in the absence of Rabi-term and the inhomogeneity, which are defined in Appendix C.1.
\begin{figure}[h]
\centering\includegraphics[width=0.9\linewidth]{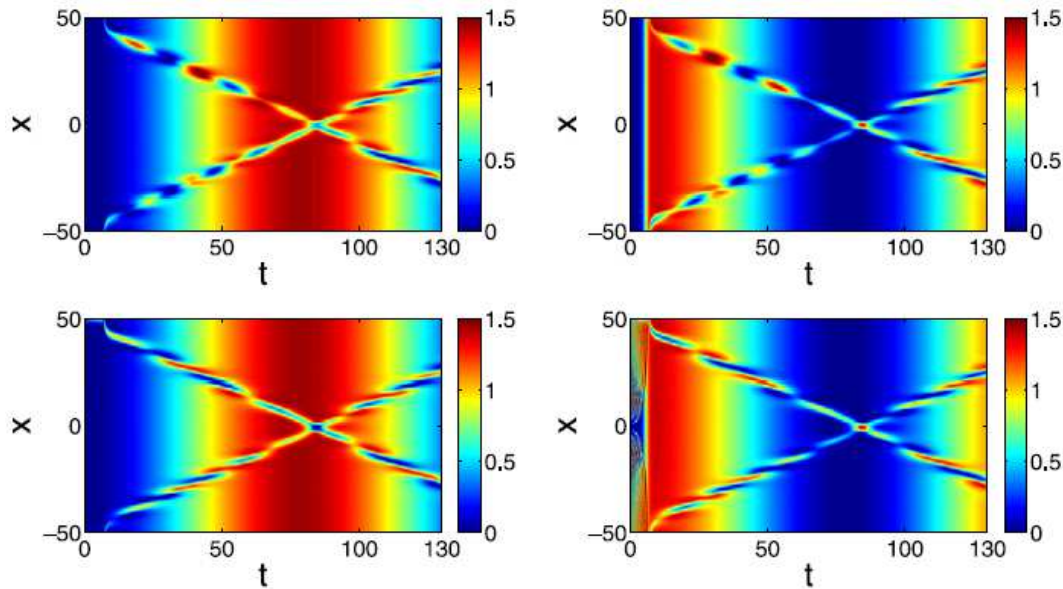}
\caption{Interaction of non-autonomous solitons in two-component GPEs for the kink-like nonlinearity
in the presence of Rabi term. Top panels show analytical results (upper left panel: $\psi_1$
component and right upper panel: $\psi_2$ component) and bottom panels show numerical results (left
bottom panel: $\psi_1$ component, and right bottom panel: $\psi_2$ component). The soliton parameters
are $k_1 = -1 + i$, $k_2 = 1 - i$, $c_1 = 3$, $b_1= 0.2$, $\alpha_1^{(1)} = \alpha_2^{(1)} = 0.03$,
$\delta= -4.5$, $\xi_1 = 0.2$, $\xi_2 = 0$, $\omega = 0.75$ and $\sigma = 0.02$.}
\end{figure}
We notice that in the absence of Rabi term $(\sigma=0)$, there is no oscillatory terms as expected. By computing the densities $(|\psi_j^{l\pm}|^2, ~j,l=1,2)$ before and after collision and noticing $|\psi_j^{l+}|^2=|\psi_j^{l-}|^2$ (since $|A_j^{l+}| = |A_j^{l-}|$, see Eqs.~(C.1)-(C.4)), we identify that the nature of collision is elastic except for a phase-shift. This phase-shift is same for both solitons $S_1$ and $S_2$ that can be found as $\frac{R_3-R_2-R_1}{2}$. There will be oscillations in bright-dark solitons and also in the background along with a modulation due to nonlinearity and Rabi term.

\subsection*{(ii) Collision dynamics in the presence of periodically modulated nonlinearity}

The collision of two solitons in the two-component GPEs with periodically modulated nonlinearity $\rho(t)$  in the absence and presence of Rabi coupling are depicted in  Figs.~12 and 13 respectively.
\begin{figure}[h]
\centering\includegraphics[width=0.9\linewidth]{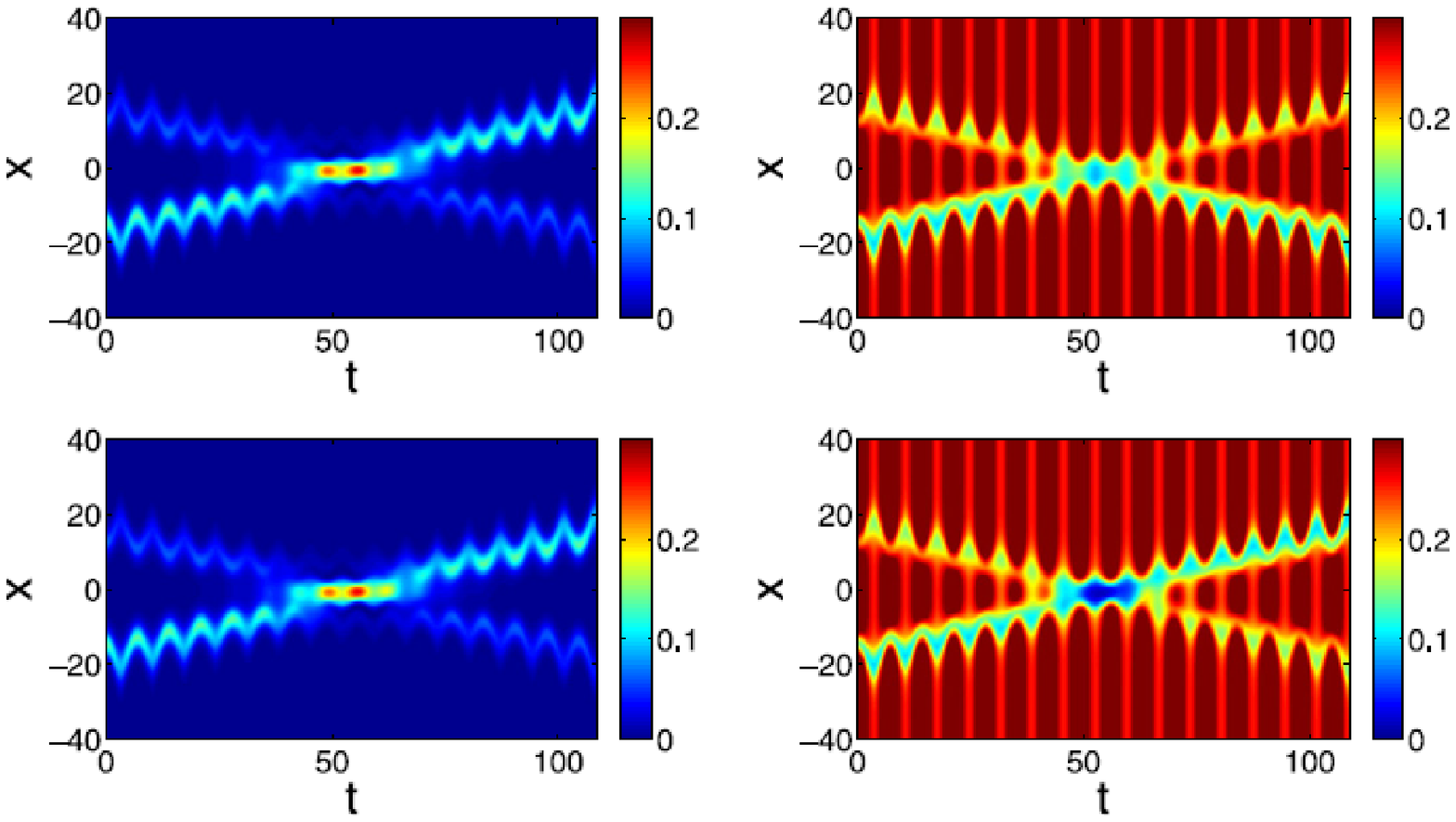}
\caption{Interaction of non-autonomous solitons in two-component GPEs for the periodically modulated
nonlinearity in the absence of Rabi term. Top panels show the exact analytical results (upper left panel:
bright component and right upper panel: dark component) and bottom panels show numerical results
(left bottom panel: bright component and right bottom panel: dark component). The parameter values
are chosen as $k_1 = -1 + i$, $k_2 = 1 - i$, $b_1= 0.2$, $c_1 = 2$, $\alpha_1^{(1)} = \alpha_2^{(1)}
= 0.02$, $\delta = 0$, $\xi_1 = 0.2$, $\xi_2 = 0$, $\varepsilon = 0.2$ and $\omega = 0.9$.}
\end{figure}
For this case the strength of the parabolic trap is given by Eq.~(\ref{perio_pot_str}).

\begin{figure}[h]
\centering\includegraphics[width=0.9\linewidth]{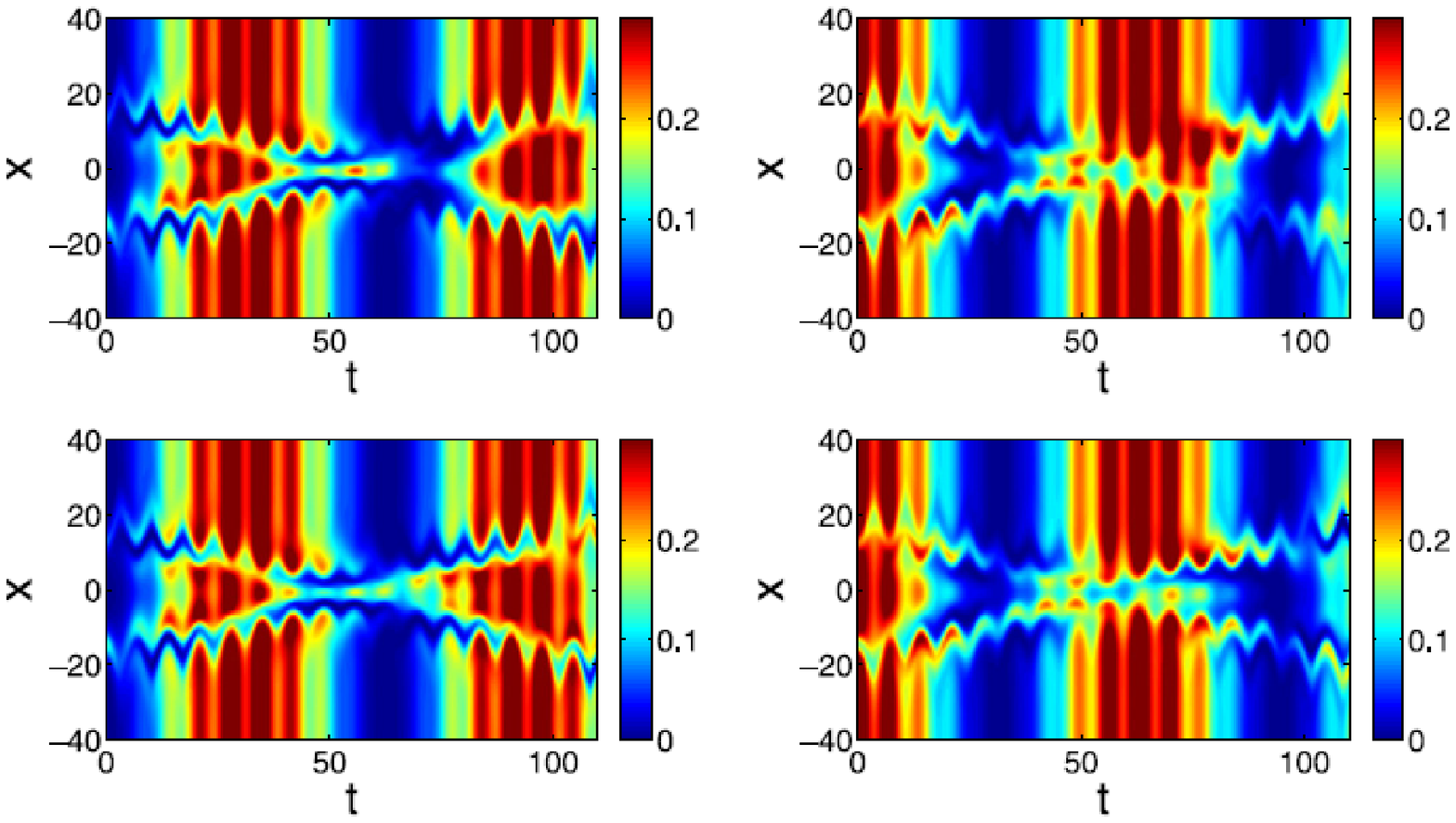}
\caption{Interaction of non-autonomous solitons in two-component GPEs for the periodically modulated
nonlinearity in presence of Rabi term. Top panels show the exact analytical results (upper left panel:
$\psi_1$ component, and right upper panel: $\psi_2$ component). Bottom panels show numerical results
(left bottom panel: $\psi_1$ component and right bottom panel: $\psi_2$ component). The parameters
are $k_1 = -1 + i$, $k_2 = 1 - i$, $b_1= 0.2$, $c_1 = 2$, $\alpha_1^{(1)} = \alpha_2^{(1)} = 0.02$,
$\delta = 0$, $\xi_1 = 0.2$, $\xi_2 = 0$, $\omega = 0.9$, $\varepsilon = 0.2$ and $\sigma = 0.05$.}
\end{figure}

The role of time modulated scattering length is to introduce periodic modulations in the soliton profile before and after collision uniformly. Meanwhile, the Rabi term leads to a periodic exchange of condensates between the components along with oscillations in the background. As in the previous example, here also the Rabi coupling makes it feasible to have both dark and bright parts of the mixed soliton in first and second components. We find that the solitons exhibit elastic collision even in the presence of Rabi coupling, though there is an exchange of atoms between the components. This example shows that the Rabi coupling does not affect the elastic nature of the collision in the two-component GPEs. An asymptotic analysis similar to that of kink-like nonlinearity can be carried out for this case too.

\section{Non-autonomous BD soliton in three-component BECs and Rabi oscillations}

Following our considerations for the binary BEC case, we now proceed with the investigation of the three-component system. The dynamics of three-component BECs in 1D with equal time-dependent interaction strengths (i.e., $g_{jl}=\rho(t)$, {\it j,l}=1,2,3), and in the presence of external time-dependent harmonic potential $V_j(x,t)(\equiv V)$, is governed by the following dimensionless non-autonomous three-coupled GPEs (see, e.g., \cite{Pana2}):
\bea
\hspace{-1.5cm}&&i \psi_{j,t}=-\frac{1}{2}\psi_{j,xx}+ \rho(t)\sum_{l=1}^3|\psi_l|^2\psi_j+\sum_{l=1, (l\neq j)}^3\sigma_l\psi_{l} +V(x,t)\psi_j,\quad j=1,2,3,
\label{3gp}
\eea
where $\psi_j$({\it x,t}) are the wave functions of the condensates. As before, coefficients
$\sigma_l$ account for the (linear) Rabi coupling and are chosen to be equal (i.e.,
$\sigma_l\equiv\sigma$,\;$l$=1,2,3); furthermore, the strength of the nonlinear coupling is given by
$\rho(t)$.

As in the two-component case, we employ a rotation transformation and a similarity transformation
(see details in Appendix~A)
and reduce Eq.~(\ref{3gp}) in the form:
\bea
iq_{j,T}+ q_{j,XX}- 2 \sum_{l=1}^3|q_l|^2q_j=0,\quad j=1,2,3.
\label{3manakov}
\eea
The above model, three-coupled NLS system (\ref{3manakov}), is also a completely integrable system (the three-component generalization of
Manakov system with defocusing nonlinearity) and the soliton solutions can be obtained by various methods, e.g.,
the Inverse Scattering Transform method, the Hirota's direct method, the B\"acklund transformation
method, etc.

\subsection{Non-autonomous mixed (bright-bright-dark) one-soliton solution}
\indent The explicit form of the mixed one-soliton solution of Eq.~(\ref{3manakov}), in the form of a
bright-bright-dark soliton, can be obtained by means of the Hirota's method \cite{{M. V},Hirota}.
Here, we consider a mixed soliton solution, with the bright part appearing in $q_1$ and $q_2$
components, and with the dark part in $q_3$ component. The exact analytical form of this
type of mixed one- and two-solitons are given in Appendix~B.2. Then we make use of the transformations, corresponding to three-component case given in Appendix.~A, for obtaining the exact non-autonomous soliton solutions of (\ref{3gp}), as in the two-component GPE system.

We again consider the same two forms for $\rho(t)$ which were discussed in the previous section during our analysis of two-component condensates.\\

\noindent{(a) \underline{\bf {Kink-like nonlinearity}}}\\

Figure~14 shows the mixed one soliton solution of the non-autonomous three-component GPEs.~(\ref{3gp}) with kink-like nonlinearity. In this case, the shape of the soliton is affected significantly by increasing the value of $\omega$ as in the two-component case.
\begin{figure}[h]
\centering\includegraphics[width=0.9\linewidth]{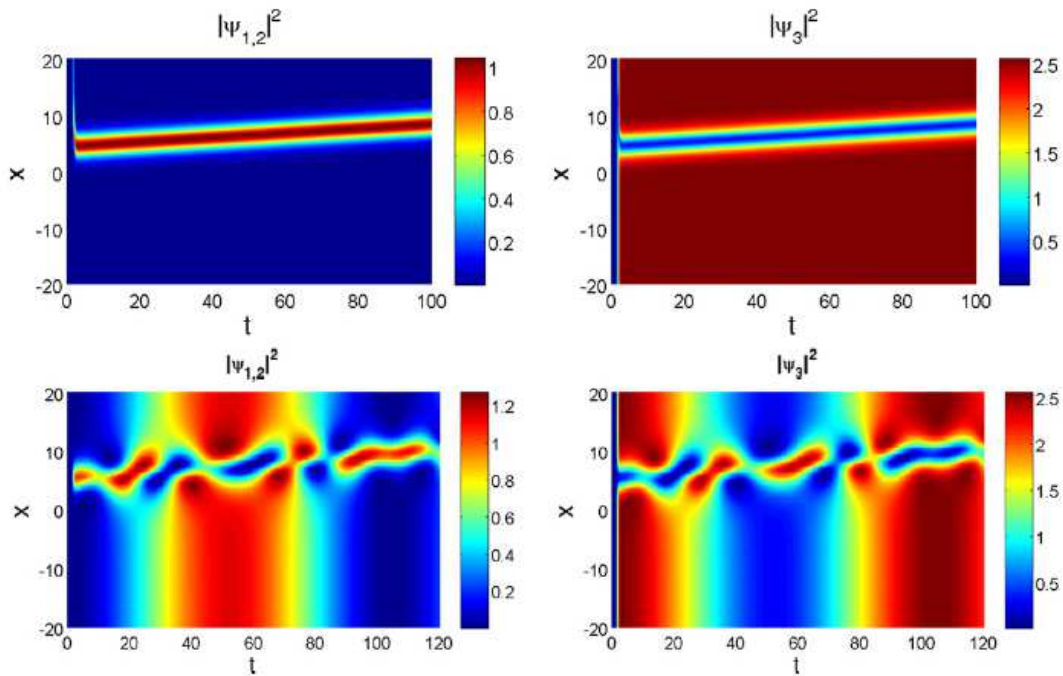}
\caption{Intensity plot of the exact three-component non-autonomous mixed one-soliton solution of Eq.~(\ref{3gp}) for the kink-like nonlinearity. Top and bottom panels show mixed non-autonomous one soliton in the absence and in the presence of Rabi coupling, respectively. The parameters are $k_1 = 0.5 + 0.02i$, $b_1 = 0.2$, $\alpha_1^{(1)} = 0.2$, $\alpha_1^{(2)} = 0.2$, $c_1 = 2$, $\delta = -5$, $\xi_1 = 0.4$, $\xi_2 = 0$, $\omega = 2.5$ and $\sigma = 0.02$.}
\end{figure}

Also, we notice that the oscillations of the background are rapid as compared to that of two component case, even for the same value of Rabi coupling $\sigma$. This can be inferred by comparing the bottom panels of Fig.~14 and Fig.~3.
Thus, an increase in the number of components can make the exchange of atoms between soliton and its background is too faster.\\

\noindent(b) \underline{{\bf {Periodically modulated nonlinearity}}}\\
The propagation of three-component non-autonomous mixed one-soliton in the presence of  periodically modulated nonlinearity having the form of Mathieu function [see Eq.~(\ref{periodic})] is shown in figure~15. As in the two component case, here also occurs oscillatory transfer of atoms among the components accompanied by oscillations of the background due to Rabi coupling.
\begin{figure}[h]
\centering\includegraphics[width=0.9\linewidth]{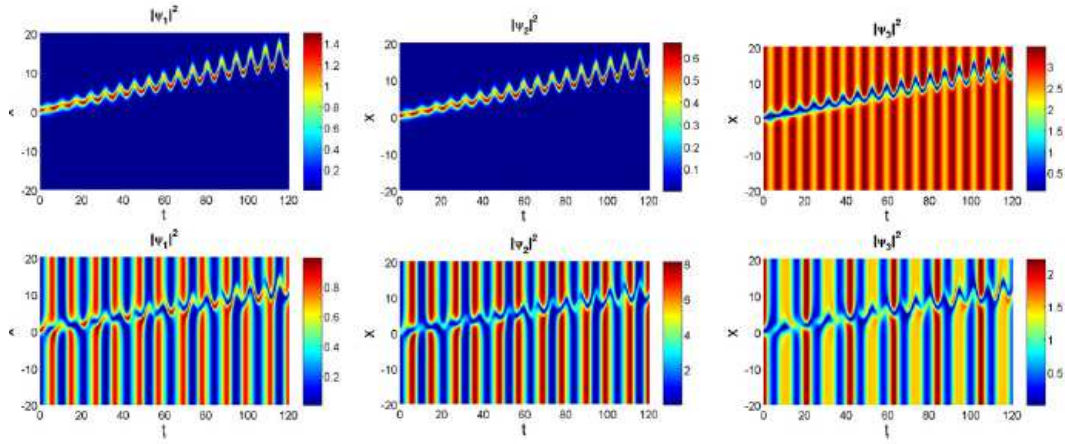}
\caption{Exact three component one-soliton solution of Eq.~(10) for the periodically modulated nonlinearity in the absence (top panels, $\psi_1$, $\psi_2$ and $\psi_3$ components) and in the presence of Rabi switch (bottom panels, $\psi_1$, $\psi_2$ and $\psi_3$ components). The parameters are chosen as $k_1 = 0.7+0.1 i$, $b_1 = 0.2$, $\alpha_1^{(1)} = 1.5$, $\alpha_1^{(2)} = 1$, $c_1 = 1$, $\xi_1 = 1$, $\xi_2 = 0$, $\omega = 0.9$, $\delta = 0$, $\varepsilon = 0.2$ and $\sigma = 0.2$.}
\end{figure}

The above two examples show that the role of Rabi coupling on three-component non-autonomous BECs is similar to that of two-component BECs except for an increase in the oscillations due to the additional component. Apart from this, we would like to give emphasis to a particular dynamical feature of the three-component system, in the presence of Rabi coupling, which is not possible in the two-component case. It can be recalled from the study on two-component non-autonomous case that it is impossible to make any component (bright/dark) of the mixed soliton to vanish completely, in the presence ---as well as in the absence--- of Rabi coupling. Contrary, in the three-component system,
it is possible to make any one of the bright parts of the mixed soliton to vanish completely before the introduction of Rabi coupling term. The additional component brings an additional arbitrariness to the non-autonomous GPE system (\ref{3gp}), which allows to make the density of any one of the bright parts of the mixed soliton to be zero before the incorporation of the Rabi term. The role of the Rabi term is to switch the condensates between all three-components. So, in the presence of Rabi term, some amount of field will be transformed to the component where there was no field before its introduction. Thus, in three-component condensates one can transfer a significant part of the condensates to a component which admits no field at all before the introduction of Rabi coupling from the other components. This is shown in Figs.~16 and 17, in which the condensate is completely absent in the second component in Fig.~16 but, after introducing the Rabi coupling, appreciable portion of condensate appears in the same component as shown in Fig.~17.

\begin{figure}[h]
\centering\includegraphics[width=0.9\linewidth]{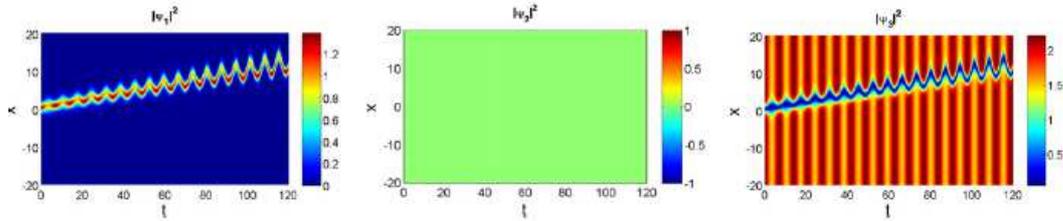}
\caption{Three-component one-soliton solution with periodically modulated nonlinearity in the absence of Rabi coupling. The parameters are $k_1 = 0.5 +0.1i$, $b_1 = 0.2$, $\alpha_1^{(1)} =1$, $\alpha_1^{(2)} = 0$, $\xi_1 = 1$, $c_1 = 1$, $\xi_2 = 0$, $\varepsilon = 0.2$ and $\omega = 0.8$.}
\end{figure}
\begin{figure}[h]
\centering\includegraphics[width=0.9\linewidth]{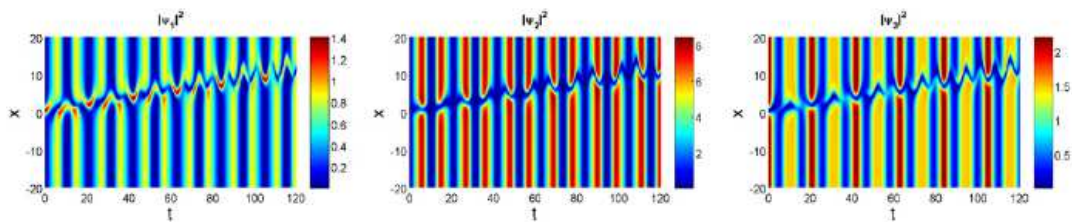}
\caption{Three-component one-soliton solution with periodically modulated nonlinearity in the presence of Rabi coupling. The parameters are $k_1 = 0.5 +0.1i$, $b_1= 0.2$, $\alpha_1^{(1)} =1$, $\alpha_1^{(2)} = 0$, $c_1 = 1$, $\xi_1 = 1$, $\xi_2 = 0$, $\varepsilon = 0.2$, $\omega = 0.8$ and $\sigma = 0.2$.}
\end{figure}

\subsection{Non-autonomous mixed (bright-bright-dark) two-soliton collision }

The explicit form of mixed two-soliton solution of (\ref{3manakov}) with the bright parts appearing
in components $(q_1,q_2)$ and the dark part in $q_3$, obtained by Hirota's bilinearization method is given
in Appendix~B.2 and the corresponding asymptotic analysis is presented in Appendix~C.2.

From the asymptotic expressions (C.5) and (C.7) one can find that the amplitudes (condensate densities) of the colliding solitons before and after collision can be related
\bea
|A_j^{l+}|^2 = |T_j^l|^2|A_j^{l-}|^2, \quad j,l=1,2, \nonumber
\eea
where $|T_j^1|^2 = \frac{e^{2\delta_{2j}-(R_3+R_2-R_1)}}{|\alpha_1^{(j)}|^2}$ and $|T_j^2|^2 = \frac{e^{2\delta_{1j}-(R_1+R_3-R_2)}}{|\alpha_2^{(j)}|^2}$, \quad $j=1,2$, where $\delta_{1j}$, $\delta_{2j}$, $R_1$, $R_2$ and $R_3$ are defined in Appendix B.2.
It is instructive to notice that $|T_j^l|$, $j,l =1,2,$ are not unimodular, in general, and also depend on dark soliton parameters $c_1$ and $b_1$ in addition to the bright soliton parameters. This will result in the energy sharing collision displaying suppression (enhancement) of condensate density in the bright part of a given soliton with commensurate changes in the bright part of the other colliding soliton. However, one can have standard elastic collision for the choice $\frac{\alpha_1^{(1)}}{\alpha_1^{(2)}}$ = $\frac{\alpha_2^{(1)}}{\alpha_2^{(2)}}$, for which $|T_j^l|$, $j,l=1,2$, become unimodular. Additionally, the colliding solitons also experience a phase-shift $\Phi_1$ = $\Phi_2$ = $\frac{(R_3-R_2-R_1)}{2}$.

It can also be inferred from the asymptotic expressions (C.6) and (C.8) that the dark solitons in the $q_3$ component always exhibit elastic collision as $|A_3^{l-}|^2=|A_3^{l+}|^2$, $l=1,2$. These dark solitons also undergo a phase-shift, same as that of bright solitons.

This study can be straightforwardly extended to integrable $N$-component CNLS system with defocusing nonlinearity, for $N>3$. It is interesting to note that the above kind of energy sharing collision in such $N$-component CNLS system can take place only if the bright part of the mixed soliton appears at least in two-components.

Such fascinating energy sharing collision in the three-component GPE system
(\ref{3manakov}), is shown in Fig.~18.
\begin{figure}[h]
\centering\includegraphics[width=0.9\linewidth]{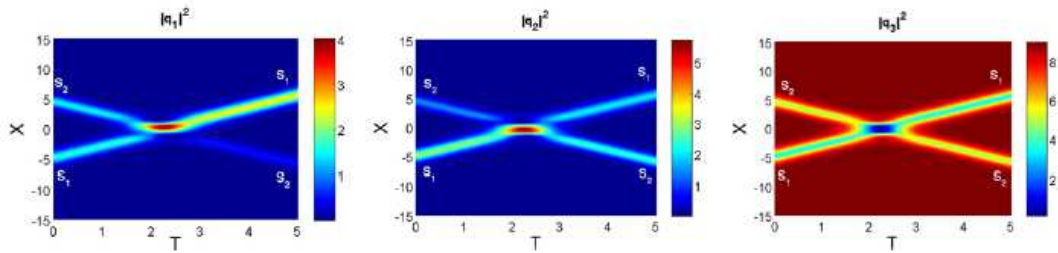}
\caption{Energy sharing collision of mixed solitons in the three-component CNLS equations (see
Eq.~(\ref{3manakov})). The parameters are $k_1 = -1 + i$, $k_2 = 1.1 - i$, $b_1= 0.2$, $c_1 = 3$,
$\alpha_1^{(1)} =0.02$ ,  $\alpha_2^{(1)} = 0.01 + 0.025i$ and
$\alpha_1^{(2)}=\alpha_2^{(2)}=0.02$.}
\end{figure}
In Fig.~18, the bright part of the mixed soliton $S_1$ is enhanced in the $q_1$ component whereas it
is suppressed in the $q_2$ component after its collision with the soliton $S_2$. The reverse scenario
takes place for the bright parts of the mixed soliton $S_2$.
Notice that the dark parts of mixed solitons $S_1$ and $S_2$ are unaffected by the collision. The
energy sharing collision is characterized by an exchange of condensates among the bright parts of the
colliding mixed solitons $S_1$ and $S_2$, leaving the dark part unaltered after collision.
This type of energy sharing collision in different integrable CNLS systems appearing in nonlinear
optics has been extensively studied in references \cite{{T. K},{M. V},{r43},{proceedmv}}.\\

The non-autonomous two-soliton solution can be expressed as
\bea
\psi_j  = \frac{1}{3}\xi_1\sqrt{2\rho}e^{i\tilde{\theta}}\left[e^{-2i\sigma t}\sum_{l=1}^3 q_l+e^{i\sigma t}\left(2q_j - \sum_{l=1,l\neq j}^3q_l\right)\right], \quad j = 1,2,3,
\label{3nonautonomous}
\eea
where $q_j$ are given in Eqs.~(B.18-B.19) in which X and T are defined by equations (A.6) and (A.7), respectively.
 Now, it is of further interest to study whether the above discussed energy sharing collision still prevails in the
presence of time-varying nonlinearity and Rabi coupling.  For illustrative purpose, again we consider
the two types of the time-dependent nonlinearities discussed in the two-component case.
\subsection*{ (i) Energy sharing collision of mixed solitons in non-autonomous three-component GPE.}

\subsection*{\bf{(a) \underline{Kink-like nonlinearity}}}

The soliton collision for a kink-like modulated nonlinearity in the context of the non-autonomous
three-component GPEs.~(\ref{3gp}), in the presence of Rabi term, is depicted in Fig.~19. The
density of the bright part of the mixed soliton $S_1$ gets enhanced and $S_2$ is suppressed after the
collision in the $\psi_1$ component, whereas the component $\psi_2$ experiences reverse effects for
$S_1$ and $S_2$, and the dark parts of mixed solitons in $\psi_3$ component remain unaltered. Thus,
the switching nature of energy sharing collision in the autonomous system (\ref{3manakov}) is
unaffected by the presence of Rabi term in the non-autonomous GP system, for this choice of
time-dependent nonlinearity and external potential. This shows that the energy sharing collision of
bright parts of the mixed solitons can exist for a wide range of time-dependence of nonlinearity,
external potential and Rabi coupling, for which the non-autonomous system (\ref{3gp}) is integrable.
An asymptotic analysis of the non-autonomous two-soliton solution (\ref{3nonautonomous}) can be carried out as in the two-component case. One can notice that the energy sharing collision still prevails as $|A_j^{l+}| \neq |A_j^{l-}|$, $j,l=1,2$, in general, which ultimately leads to $|\psi_j^{l+}|$ $\neq$ $|\psi_j^{l-}|$.
The phase-shift can be found to be the same for both the solitons and is given by $\frac{(R_3-R_2-R_1)}{2}$, where $R_1$, $R_2$ and $R_3$ are defined in Appendix~B.

In the two-component case, we observe a gradual increase in the density of condensates in the bright
parts in the region $t$ = $2$ to $6$ and it remains constant for $t>6$, due to the nature of
nonlinearity (Fig.~10) and onset of Rabi oscillations due to Rabi coupling (see Fig.~11). Note that,
there the collision is elastic in the absence as well as in the presence of Rabi coupling and
is not an energy sharing one. Thus, in two-component BECs the growth of condensate is purely due to
the nature of $\rho(t)$.
\begin{figure}[h]
\centering\includegraphics[width=0.9\linewidth]{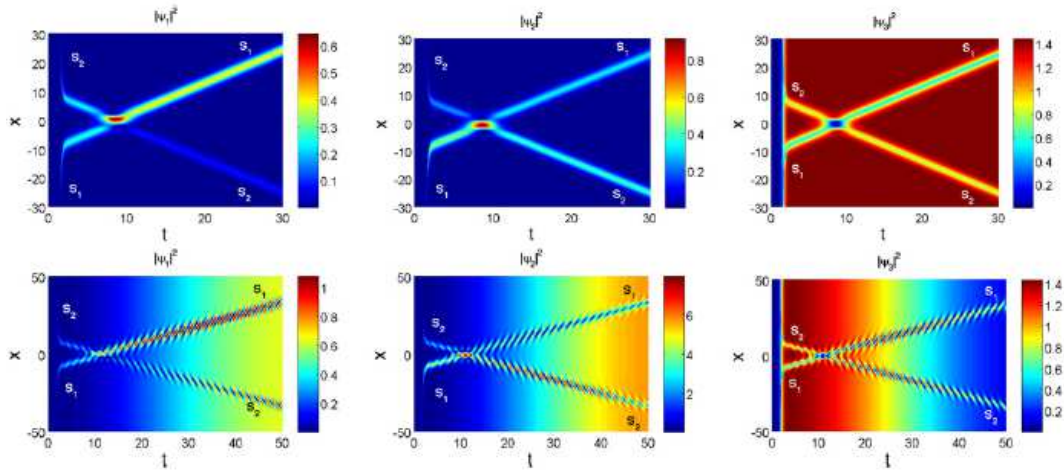}
\caption{Energy sharing collision of mixed solitons in the non-autonomous three-component
GP Eq.~(\ref{3gp}) with kink like nonlinearity in the absence (top panels) and in the presence
(bottom panel) of Rabi coupling. The soliton parameters are $k_1 = -1 + i$, $k_2 = 1.1 - i$, $b_1= 0.2$, $c_1 = 3$,
$\alpha_1^{(1)} = 0.02$, $\alpha_2^{(1)} = 0.01 + 0.025 i$, $\alpha_1^{(2)}=\alpha_2^{(2)}=0.02$,
$\xi_1 = 0.2$, $\xi_2 = 0$, $\delta = -4$, $\omega = 2$ and $\sigma = 0.02$.}
\end{figure}
However, in the present case, for the mixed soliton $S_2$ there is a suppression in the density of
its bright part in $\psi_1$ after interaction due to the energy sharing collision. This suppression
balances the enhancement in its density in this regime due to the nature of the time-varying
nonlinearity, and will result in standard soliton $(S_2)$ with constant amplitude after interaction
(see Fig.~19).
Also, the Rabi term leads to beating effects in this soliton $S_2$ in $\psi_1$, but with a smaller
period as compared to that of the two-component case. However, the opposite effect takes place for
$S_2$ in the second component $\psi_2$. This shows that energy sharing collision can be profitably
used in altering the exchange of condensates through Rabi switching.

\subsection*{\bf{(b) \underline{Periodic modulated nonlinearity}}}

Energy sharing collision of mixed solitons in the three-component GPEs with periodically modulated
nonlinearity is shown in Fig.~20. We observe that in this case also the energy sharing collision for
the non-autonomous system (\ref{twogp}) takes place as that of the autonomous system (\ref{2manakov})
(see Fig.~18).
\begin{figure}[h]
\centering\includegraphics[width=0.9\linewidth]{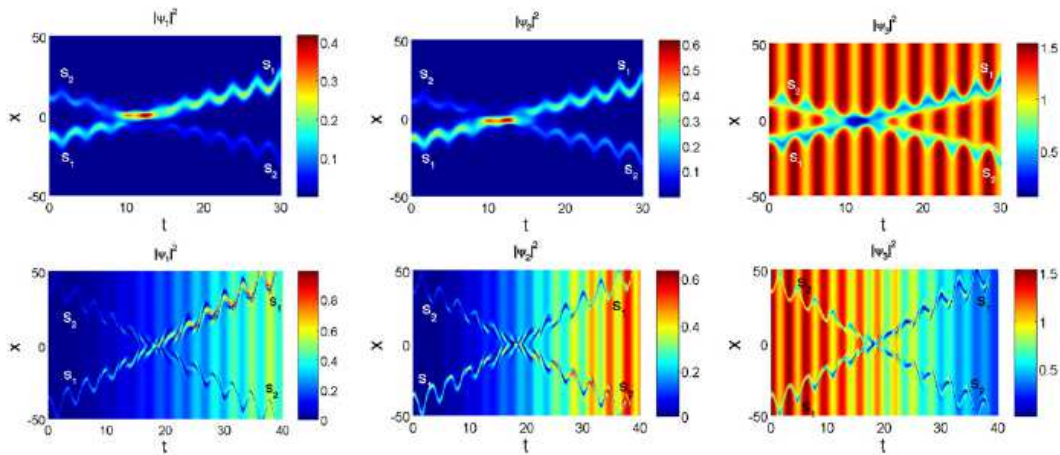}
\caption{Energy sharing collision of mixed soliton in the non-autonomous three-component GPEs with
periodically modulated nonlinearity in the absence (top panels) and in the presence (bottom panels)
of Rabi coupling. The choice of soliton parameters are $k_1 = -1 + i$, $k_2 = 1- i$, $b_1 = 0.2$,
$c_1= 2$, $\alpha_1^{(1)} = 0.02$, $\alpha_2^{(1)} = 0.01 + 0.025 i$, $\alpha_1^{(2)}=\alpha_2^{(2)}=
0.02$, $\delta = 0$, $\xi_1 = 0.4$, $\xi_2 = 0$, $\omega = 2$, $\varepsilon = 0.2$ and $\sigma= 0.02$.}
\end{figure}

We observe another dramatic switching of condensates during collision of two solitons in the
non-autonomous three-component GPEs.~(\ref{3gp}) with periodically modulated time-dependent
nonlinearity, for smaller values of Rabi term $\sigma$. Here, before collision, we have an
oscillating bright soliton part $(S_2)$ in the $\psi_1$ component, which completely transforms to an
oscillating mixed~(bright-dark) soliton in the same component after collision. In the second
component, $\psi_2$, the amplitude of bright soliton part $(S_2)$ gets enhanced after the collision.
For the soliton $S_1$, in $\psi_1$ $(\psi_2)$ component enhancement (suppression) with
periodic oscillations takes place. In the component $\psi_3$ the solitons undergo elastic collision
and there occurs only switching of condensates due to Rabi coupling. Notably, in $\psi_3$ component
the pure bright part of $S_1$ is transformed to mixed (BD) part after collision (see $\psi_3$
component in the bottom panel of Fig.~20). This interesting collision is a consequence of the
combined effects of time-dependent nonlinearity and external potential, Rabi term and the exchange of
condensates due to energy sharing collision. Another noticeable effect arising due to the Rabi
coupling, particularly for this type of periodic nonlinearity, is an increase in the relative
separation distance between the solitons well before and after collision. Furthermore, the collision
takes place faster in the presence of Rabi term.

\section{Conclusions}

In this work, we have studied the dynamics of non-autonomous mixed bright-dark matter-wave solitons
in two- and three-component Bose-Einstein condensates. Our setting included a time-dependent
parabolic potential and scattering length, as well as Rabi coupling between separate hyperfine
states. We have transformed the non-autonomous two- and three-component Gross-Pitaevskii equations, into a set of integrable defocusing two- and three-component Manakov autonomous systems by
means of two successive transformations. These transformations
can be viewed as a rotation followed by a similarity transformation. Then, with the aid of the two-
and three-component soliton solutions of defocusing Manakov systems, and by inverting the
transformations, we have studied the dynamics of single- and multiple-non-autonomous matter-wave
solitons. Our considerations involved two different choices of nonlinearities namely, a kink-like and
Mathieu-like ones, together with their corresponding time-varying harmonic potentials.

Our study on the two-component case has shown that the nature of soliton propagation depends
significantly on the nature of nonlinearity and the associated potential. The switching depends
purely on the Rabi coupling. Here, the bright-dark matter-wave solitons exhibit elastic collisions
along with oscillations due to Rabi coupling, and profile modulation due to the nonlinearity
$\rho(t)$ and strength of the potential. In this two-component non-autonomous case, the dark-bright
solitons exist as ``symbiotic'' structures, i.e., the bright component exist only due to the presence
of its dark counterpart. Hence  it is impossible to make any part (either bright or dark) to be zero.
Also, Rabi coupling makes it possible to have a special type of mixed soliton in which bright and
dark parts can co-exist
in the same component. Such structures are reminiscent to the so-called beating dark-dark solitons,
which have recently been observed in autonomous binary BECs \cite{engels2,pe5}. The present study
shows the possibility of observing such structures, with co-existing bright and dark parts in the
same component, also in non-autonomous multi-component systems in the presence of Rabi coupling. Our
analytical results of the two-component non-autonomous GPE system are found to be in very good
agreement with results of direct numerical simulations.

We have also studied non-autonomous three-component BECs. An important observation in this case is
that the bright part of the mixed soliton can be absent in any of the three-component condensate in
the absence of Rabi coupling. The introduction of Rabi coupling makes it feasible to have some
portion of mixed-soliton in that component. Also, in the three-component BECs the number of
oscillations increases as compared to that of two-component case and ultimately the exchange of atoms
between the soliton and background also increases in a periodic manner.

Finally, we considered the collision of non-autonomous solitons in the three-component condensate. In
this case, the non-autonomous matter-wave solitons undergo interesting energy-sharing collision,
leading to an exchange of atoms, in addition to Rabi coupling, in the bright ($\psi_1$ and $\psi_2$)
components; notice that in the third component ($\psi_3$) the solitons undergo elastic collision. The
matter-wave mixed (bright-bright-dark) soliton collision in non-autonomous three-component
condensates also displays another fascinating feature, namely the reversal of the type of mixed-
soliton part (i.e., bright type to dark type or vice-versa).

Our study can be extended to other types of forms for nonlinearity modulation by identifying the
corresponding modulation of the harmonic potential using the Riccati Eq.~(\ref{ricatti}) and vice-
versa. The agreement between the numerical and analytical results indicates that the presented exact
non-autonomous solution can well be utilised for the purpose of studying non-integrable multi-
component GPE type systems with Rabi coupling. It is of future interest to investigate the effect of
spatial and spatio-temporal modulations of nonlinearity. Works along these directions are in
progress. This study may have ramifications in the design of matter-wave devices (e.g., switches) and
also in experiments involving growth and oscillations of condensates.

\section*{Acknowledgments}

The work of T.K. is supported by Department of Science and Technology (DST), Government of India, in
the form of a major research project. R.B.M. acknowledges the financial support from DST in the form
of Project Assistant. T.K. and R.B.M. also thank the principal and management of Bishop Heber College
for constant support and encouragement. The work of F.T., H.E.N., and D.J.F. was partially supported
by the Special Account for Research Grants of the University of Athens.

\appendix
\section{Mapping non-autonomous GPEs to integrable CNLS systems}
\setcounter{section}{1}
\label{a1}

In this Appendix, we explicitly show that the non-autonomous two- and three-component GP systems
[cf. Eqs.~(\ref{twogp}) and (\ref{3gp})]
with Rabi coupling can be transformed to integrable CNLS systems, i.e., the Manakov model and its three component generalization, with defocusing nonlinearities; this will be done by means
of two successive transformations. In particular, in the case of the two-component BEC,
first we apply the following unitary transformation to Eq.~(1) with $g_{jl} = \rho(t)$ and $\sigma_1 = \sigma_2=\sigma$:
\bea
\left(
  \begin{array}{c}
    \psi_1 \\
    \psi_2 \\
  \end{array}
\right)= \left(
  \begin{array}{cc}
    \mbox{cos}(\sigma t) & -i \mbox{sin}(\sigma t)\\
    -i  \mbox{sin}(\sigma t) &  \mbox{cos}(\sigma t) \\
  \end{array}
\right)
\left(
  \begin{array}{c}
    \phi_1 \\
    \phi_2 \\
  \end{array}
\right).
\label{rabi}
\eea
In the case of the three-component BECs, a similar transformation is applied to Eq.~(\ref{3gp});
this transformation is of the form \cite{r35,{B. D}}:
\bea
\hspace{-2.5cm}
\left(
  \begin{array}{c}
    \psi_1 \\
    \psi_2 \\
    \psi_3 \\
  \end{array}
\right)= \frac{1}{3}\left(
  \begin{array}{ccc}
    (2e^{i \sigma t}+e^{-2i\sigma t}) & (e^{-2i\sigma t}- e^{i \sigma t}) & (e^{-2i\sigma t}- e^{i \sigma t})\\
    (e^{-2i\sigma t}- e^{i \sigma t}) & (2e^{i \sigma t}+ e^{-2i\sigma t})& (e^{-2i\sigma t}- e^{i \sigma t})\\
    (e^{-2i\sigma t}- e^{i \sigma t}) & (e^{-2i\sigma t}- e^{i \sigma t}) & (2e^{i \sigma t}+ e^{-2i\sigma t})\\
  \end{array}
\right)
\left(
  \begin{array}{c}
    \phi_1 \\
    \phi_2 \\
    \phi_3
  \end{array}
\right).
\nonumber \\
\label{tut}
\eea
This way, we obtain the following set of non-autonomous equations without the Rabi coupling term:
\bea
i \phi_{j,t}=-\frac{1}{2}\phi_{j,xx}+\left(\rho(t) \sum_{l=1}^N |\phi_l|^2+ V(x,t)\right)\phi_j,
\label{autonomous}
\eea
where $N=2$ and $j=1,2,$ for Eq.~(1), while $N=3$ and $j=1,2,3$ for Eq.~(\ref{3gp}).
Then, by performing the similarity transformation
\bea
&&\phi_j(x,t) = \xi_1\sqrt{2\rho(t)}~e^{i\tilde{\theta}}q_j(X,T),
\label{similarity}
\eea
where
\bea
&&\tilde{\theta} = -\frac{1}{2}\left[\frac{d}{dt}(\ln \rho)\right]x^2 + 2 \xi_2 \xi_1^2 \left(\rho x - \xi_2 \xi_1^2\int_0^t \rho^2 dt\right),\\
&&X = \sqrt {2}~\xi_1 \left(\rho x - 2\xi_2 \xi_1^2\int_0^t \rho^2 dt\right),\\
&&T = \xi_1^2 \int_0^t \rho^2 dt,
\eea
\label{eqn10}
with $\xi_1$ and $\xi_2$ being arbitrary real constants, Eq.~(\ref{autonomous}) can be transformed
into the system of Eq.~(\ref{2manakov}) for the two-component system, or to
Eq.~(\ref{3manakov}) for the three-component system,
%
with the condition
\bea
\frac{d\Lambda}{dt} - \Lambda^2 - \Omega^2(t)= 0,
\label{ricatti}
\eea\ees
where $\Lambda = (\rho_t/\rho)$. Equation~(\ref{ricatti}) is nothing but a Riccati-type equation. Notice that similar type of transformation has been reported for two-component BECs \cite{rajend}
but in the absence of Rabi coupling.


\appendix
\setcounter{section}{1}
\section{One- and two-soliton solutions of the integrable two- and three- component defocusing CNLS systems}
\label{b2}

\subsection{Bright-dark solitons in the two-component system (\ref{2manakov})}

The bright-dark one-soliton solution of Eq.~(\ref{2manakov}), with bright (dark) part appearing in the $q_1$ ($q_2$) component, obtained by Hirota's bilinearization method \cite{Shepp,proceedmv}.
\bea
\hspace{-1.5cm}
&&q_{1}(X,T) = \sqrt{|c_{1}|^2 cos^2 \varphi_{1}- k^2_{1R}}~\mbox{sech}[k_{1R}(X- 2k_{1I}T)+R/2]~e^{i(\eta_{1I}+ \theta)},
\label{1solXT1} \\
\hspace{-1.5cm}&&q_{2}(X,T) = -c_{1}~e^{i(\zeta_1+\varphi_1)}~(\mbox{cos}\varphi_1~\mbox{tanh}[k_{1R}(X-2k_{1I}T)+R/2]+ i~ \mbox{sin}\varphi_1),
\label{1solXT2}
\eea
where $\varphi_1 = \mbox{tan}^{-1}(\frac{k_{1I}-b_1}{k_{1R}})$,
$\zeta_1$ = -$(b_1^2 + 2|c_1|^2)T + b_1 X$,
$\alpha_1^{(1)} = \alpha_{1R}^{(1)} + i \alpha_{1I}^{(1)}$,
$\theta = \mbox{tan}^{-1}(\frac{\alpha_{1I}}{\alpha_{1R}})$, $\eta_1 = k_1X +i(k_1^2-2|c_1|^2)T$ and
$e^{R} = -\frac{1}{(k_1+k_1^*)^2}\left(1-\frac{|c_1|^2}{|k_1-ib_1|^2}\right)^{-1}$. The bright part
of the mixed soliton is characterized by three complex parameters $k_1$, $\alpha_1$, and $c_1$, and
one real parameter $b_1$. Note that $\alpha_1$ does not affect the amplitude of the soliton. The
dark soliton part of the mixed soliton also influences the bright part through the parameters $c_1$
and $b_1$. Also, the solution becomes non-singular only for the choice $|c_1|^2>|k_1 - i b_1|^2$. As
a consequence of this, it is
impossible to make either one of the soliton part (i.e., dark/bright) completely zero. Thus, the
bright and the dark parts of the mixed soliton co-exist but appear in separate components due to the
presence of the other and can be viewed as ``symbiotic solitons'', as mentioned in the introduction.

On the other hand, the explicit form of mixed two-soliton solution of the integrable
2-CNLS system (\ref{2manakov}) can be written as \cite{Shepp}:
\bea
\!\!\!\!\!\!\!\!\!\!\!\!\!\!\!\!\!\!\!\!
&q_{1}(X,T) = &\frac{1}{D}\left(\alpha_1^{(1)}e^{\eta_1}+\alpha_2^{(1)}e^{\eta_2} + e^{\eta_1 + \eta_1 ^* + \eta_2 + \delta_{11}} + e^{\eta_2 + \eta_2 ^* + \eta_1 + \delta_{21}}\right),\\
\!\!\!\!\!\!\!\!\!\!\!\!\!\!\!\!\!\!\!\!
&q_{2}(X,T) = &\frac{c_1e^{i\zeta_1}}{D}\left[1 + e^{\eta_1 + \eta_1 ^* + Q_{11}}+ e^{\eta_1 + \eta_2 ^* + Q_{12}}+ e^{\eta_2 + \eta_1 ^* + Q_{21}}\right.\nonumber\\ &&\left.~~~~~~~~+ e^{\eta_2 + \eta_2 ^* + Q_{22}}+ e^{\eta_1 + \eta_1 ^* + \eta_2+ \eta_2 ^* + Q_{3}} \right],
\label{2c2sol}
\eea
where
\bea
\hspace{-1.5cm}&&D = 1 + e^{\eta_1 + \eta_1 ^* + R_1} + e^{\eta_1 + \eta_2 ^* + \delta_0} + e^{\eta_2 + \eta_1 ^* + \delta_0^*} + e^{\eta_2 + \eta_2 ^* + R_2}+ e^{\eta_1 + \eta_1 ^* +\eta_2 + \eta_2 ^*+ R_3}.
\label{2c21sol}
\eea

The various parameters in the above equation are defined below:
\bea
\!\!\!\!\!\!\!\!\!\!\!\!\!\!
&& e^{R_1} = \mu_{11},~e^{R_2} = \mu_{22},~ e^{\delta_0}= \mu_{12},~ e^{\delta_0^\ast}=
\mu_{21},~\zeta_1 = -(b_1^2 + \lambda)T + b_1 X,
\\
\!\!\!\!\!\!\!\!\!\!\!\!\!\!
&&\eta_j = k_j X+ i(k_j^2 - \lambda)T,~\lambda = 2|c_1|^2, e^{Q_{ij}} = -\left(\frac{k_i - i b_1}
{k_j^* + i b_1}\right)\mu_{ij},~i,j=1,2,
\\
\!\!\!\!\!\!\!\!\!\!\!\!\!\!
&&e^{Q_3} = - \left(\frac{(k_1 - i b_1)(k_2 - i b_1)}{(k_1^* + i b_1)(k_2^* + i b_1)}\right)e^{R_3},
\\
\!\!\!\!\!\!\!\!\!\!\!\!\!\!
&&e^{R_3} = |k_1 - k_2|^2 \mu_{11}\mu_{12}\mu_{21}\mu_{22}(\chi_{12}\chi_{21} -\chi_{11}\chi_{22}),
\\
\!\!\!\!\!\!\!\!\!\!\!\!\!\!
&&\mu_{il} = \frac{1}{(k_i + k_l^*)\chi_{il}},\quad i,l=1,2,
\\
\!\!\!\!\!\!\!\!\!\!\!\!\!\!
&& e^{\delta_{11}}= (k_2-k_1)\mu_{11}\mu_{21}(\alpha_2^{(1)}\chi_{21}- \alpha_1^{(1)}\chi_{11}),\\
\!\!\!\!\!\!\!\!\!\!\!\!\!\!
&& e^{\delta_{21}}= (k_2-k_1)\mu_{12}\mu_{22}(\alpha_2^{(1)}\chi_{22}-
\alpha_1^{(1)}\chi_{12}),
\\
\!\!\!\!\!\!\!\!\!\!\!\!\!\!
&&\chi_{il} = -\left[\frac{(k_i + k_l^*)}{(\alpha_i^{(1)}\alpha_l^{(1)*})} \left(1 - \frac{|c_1|^2}{(k_i -ib_1)(k_l ^* +  ib_1)}\right)\right],\quad i,l=1,2.
\eea

\subsection{Bright-bright-dark solitons in the three-component system (\ref{3manakov})}

The one-soliton solution of Eq.~(\ref{3manakov}), in the form of a
bright-bright-dark soliton, obtained by Hirota's bilinearization method is given below:
\bea
\!\!\!\!\!\!\!\!\!\!\!\!\!\!\!\!\!\!\!\!\!\!\!\!\!\!\!\!\!\!
&&q_{j}(X,T) = A_j\sqrt{|c_{1}|^2 cos^2 \varphi_{1}- k^2_{1R}}~\mbox{sech}[k_{1R}(X- 2k_{1I}T)+ R/2]e^{i\eta_{1I}}, ~~j=1,2,\\
\!\!\!\!\!\!\!\!\!\!\!\!\!\!\!\!\!\!\!\!\!\!\!\!\!\!\!\!\!\!
&&q_{3}(X,T) = -c_{1} e^{i(\zeta_1+\varphi_1)}~(\mbox{cos}\varphi_1\mbox{tanh}[k_{1R}(X-2k_{1I}T)+R/2]+ i\mbox{sin}\varphi_1),
\label{3c1sol}
\eea
where
\bea
&&\varphi_1 = \mbox{tan}^{-1}\left(\frac{k_{1I}-b_1}{k_{1R}}\right),~
A_j = \left(\frac{\alpha_1^{(j)}}{\sqrt{|\alpha_1^{(1)}|^2+|\alpha_1^{(2)}|^2}}\right), \quad j=1,2,\\
&&e^{R} = \frac{\sum_{j=1}^2(\alpha_1^{(j)} \alpha_1^{(j)^*})}{(k_1+k_1^*)^2}\left(\frac{|c_1|^2}{|k_1-ib_1|^2}-1\right)^{-1}. \
 \eea
The above soliton solution is characterized by three complex parameters $c_1$, $\alpha_1^{(1)}$, $\alpha_1^{(2)}$, and one real parameter $b_1$ along with the non-singular condition $|c_1|^2>|k_1 - i b_1|^2$. Parameters $A_j$ may be viewed as spin-polarization in the case of spinor condensates \cite{{J. I},{r1}}.

On the other hand, the respective two-soliton solution of Eq.~(\ref{3manakov}) can be obtained as
follows:
\bea
\hspace{-2.5cm}
\!\!\!\!\!\!\!\!\!\!\!\!\!\!\!\!\!\!\!\!\!\!\!\!\!\!\!\!\!\!
&q_{j}(X,T) =&\frac{1}{D}\left(\alpha_1^{(j)}e^{\eta_1}+\alpha_2^{(j)}e^{\eta_2} + e^{\eta_1 + \eta_1 ^* + \eta_2 + \delta_{1j}} + e^{\eta_2 + \eta_2 ^* + \eta_1 + \delta_{2j}}\right),\quad j=1,2,\\
\hspace{-2.5cm}
\!\!\!\!\!\!\!\!\!\!\!\!\!\!\!\!\!\!\!\!\!\!\!\!\!\!\!\!\!\!
&q_{3}(X,T) =&\frac{c_1e^{i\zeta_1}}{D}\left[1 + e^{\eta_1 + \eta_1 ^* + Q_{11}}+ e^{\eta_1 + \eta_2 ^* + Q_{12}}+ e^{\eta_2 + \eta_1 ^* + Q_{21}}\right.\nonumber\\
\hspace{-1.5cm}&&\left.~~~~~~~~+ e^{\eta_2 + \eta_2 ^* + Q_{22}}+ e^{\eta_1 + \eta_1 ^* + \eta_2+ \eta_2 ^* + Q_{3}} \right],
\label{3manakovsolution}
\eea
where
\bea
\hspace{-1.5cm}&&D = 1 + e^{\eta_1 + \eta_1 ^* + R_1} + e^{\eta_1 + \eta_2 ^* + \delta_0} + e^{\eta_2 + \eta_1 ^* + \delta_0^*} + e^{\eta_2 + \eta_2 ^* + R_2}+ e^{\eta_1 + \eta_1 ^* +\eta_2 + \eta_2 ^*+ R_3}.
\eea
The various parameters in the above equation are defined below
\bea
&& \zeta_1 = -(b_1^2 +\lambda)T + b_1 X,\\
&&e^{\delta_{1j}}= (k_2-k_1)\mu_{11}\mu_{21}(\alpha_2^{(j)}\chi_{21}- \alpha_1^{(j)}\chi_{11}), \\
&&e^{\delta_{2j}}= (k_2-k_1)\mu_{12}\mu_{22}(\alpha_2^{(j)}\chi_{22}- \alpha_1^{(j)}\chi_{12}), \quad j = 1,2,\\
&&\mu_{il} = \frac{1}{(k_i + k_l^*)\chi_{il}},\\
&&\chi_{il} = -\left[\frac{(k_i + k_l^*)}{\sum_{j=1}^2(\alpha_i^{(j)}\alpha_l^{(j)*})} \left(1 - \frac{|c_1|^2}{(k_i -ib_1)(k_l ^* +  ib_1)}\right)\right],\quad i,l=1,2.
\eea
The other parameters $R_1$, $R_2$, $R_3$, $\delta_0$, and $\delta_0^{*}$ in the above equation are similar to that of the mixed two-soliton solution for the two-component case, given after Eq.~(\ref{2c21sol}), with the above redefinition of $\chi_{il}$.

\appendix
\setcounter{section}{2}
\section{Asymptotic analysis}
\label{c1}
\subsection{Asymptotic analysis of two-soliton solution of the Manakov system (\ref{2manakov})}
In this Appendix, we present the asymptotic forms of the two colliding solitons obtained by performing the asymptotic analysis of the two-soliton solution given by Eqs.~(B.3)-(B.5). For the analysis, we choose $k_{1R}<k_{2R}$ and $k_{1I}>k_{2I},$ without loss of generality.

\noindent{\bf{Before collision}}
\bea
q_1^{l-}=A_1^{l-}\mbox{sech}\left(\eta_{lR}+\frac{R_l}{2}\right)e^{i\eta_{lI}},\\
q_2^{l-}=A_2^{l-}\left[\mbox{cos}\varphi_l\mbox{tanh}\left(\eta_{lR}+\frac{R_l}{2}\right)+i\mbox{sin}\varphi_l\right], \quad l=1,2.
\eea
{\bf{After collision}}
\bea
q_1^{l+}=A_1^{l+}\mbox{sech}\left( \eta_{lR}+\frac{R_3-R_{3-l}}{2}\right)e^{i\eta_{lI}},\\
q_2^{l+}=A_2^{l+}\left[\mbox{cos}\varphi_l\mbox{tanh}\left(\eta_{lR}+\frac{R_3-R_{3-l}}{2}\right)+i\mbox{sin}\varphi_l\right], \quad l=1,2,
\eea
where $\eta_{lR}= k_{lR}(X-2k_{lI}T)$, $\eta_{lI}=k_{lI}X-(k_{lR}^2-k_{lI}^2-2|c_1|^2)T$,
$A_1^{l-}=\frac{\alpha_l^{(1)}}{2}e^{-R_l/2}$ , $A_2^{l-}=-c_1e^{i(\zeta_1+\varphi_l)}$,
$A_1^{l+}=\frac{1}{2}e^{-(R_{3-l}+R_3)/2+\delta_{l1}}$ and $A_2^{l+}=c_1e^{-i(\zeta_1+2\varphi_{3-l}+\varphi_l)}$.
Here and in the following, as mentioned in the text, the superscript (subscript) in $q$ (or $\psi$) and in $A$ represents the number of soliton (component) while -(+) appearing in the corresponding quantities indicates their form before (after) collision. All the quantities found in Eqs.~(C.1)-(C.4) are defined below Eq.~(B.5). From the above expressions it can be easily verified that $|A_i^{l+}|^2$ = $|A_i^{l-}|^2$, $i,l=1,2$.

\subsection{Asymptotic analysis of two-soliton solution of integrable three coupled NLS system (\ref{3manakov})}
In this Appendix, we present the results of the asymptotic analysis of two-soliton solution of Eq.~(\ref{3manakov}) [see Eqs.~(B.18) and (B.19)] briefly. As before, here also without loss of generality, we choose $k_{1R}<k_{2R}$, $k_{1I}>k_{2I}$. The asymptotic forms are given below:

\noindent {\bf{Before collision}}
\bea
q_j^{l-}&=&A_j^{l-}\mbox{sech}\left( \eta_{lR}+\frac{R_l}{2}\right)e^{i\eta_{lI}},\\
q_3^{l-}&= &A_3^{l-}\left[\mbox{cos}\varphi_l\mbox{tanh}\left(\eta_{lR}+\frac{R_l}{2}\right)+i\mbox{sin}\varphi_l\right], \quad j,l =1,2,
\eea

\noindent {\bf{After collision}}
\bea
q_j^{l+} &= &A_j^{l+}\mbox{sech}\left( \eta_{lR}+\frac{R_3-R_{3-l}}{2}\right)e^{i\eta_{lI}}, \\
q_3^{l+} &= &A_3^{l+}\left[\mbox{cos}\varphi_l\mbox{tanh}\left(\eta_{lR}+\frac{R_3-R_{3-l}}{2}\right)+i\mbox{sin}\varphi_l\right], \quad j,l =1,2,
\eea
where $\eta_{lR} = k_{lR}(X-2k_{lI}T)$, $\eta_{lI}=k_{lI}X-(k_{lR}^2-k_{lI}^2-2|c_1|^2)T$,
$A_j^{l-}=\frac{\alpha_l^{(j)}}{2}e^{-\frac{R_l}{2}}$, $A_3^{l-} = -c_1 e^{i(\zeta_1+\varphi_l)}$,
 $A_j^{l+}=\frac{1}{2}e^{-(R_{3-l}+R_3)/2+\delta_{lj}}$ and $A_3^{l+} = c_1 e^{i(\zeta_1+2\varphi_{3-l}+\varphi_l)}$. The quantities $R_1$, $R_2$, $R_3$ and $\delta_{lj}$, $l,j =1,2$, appearing in the above expressions are defined below Eq.~(B.5) and $\varphi_l = \mbox{tan}^{-1}\left(\frac{k_{lI}-b_1}{k_{lR}}\right)$, $l=1,2$.

\end{document}